\documentclass[twocolumn,secnumarabic,footinbib,tightenlines,nobibnotes,aps,prb,unsortedaddress,superscriptaddress]{revtex4-1}
\usepackage[usenames,dvipsnames]{color}
\usepackage{mathtools}
\usepackage{hyperref}
\usepackage{graphicx}
\usepackage[tight]{subfigure}
\usepackage{bbding}
\usepackage{epstopdf}
\usepackage{graphicx}
\usepackage{amsmath}
\usepackage{amsfonts}
\usepackage{amssymb}
\usepackage{bezier} 
\usepackage{bm}
\usepackage{multirow}

\newcommand{\pmu}{\partial_\mu}
\newcommand{\pnu}{\partial_\nu}
\newcommand{\avbra}[1]{\blangle #1 \brangle}

\DeclareMathOperator{\Imag}{Im}
\DeclareMathOperator{\Real}{Re}
\newcommand{\mib}{\bm}
\newcommand{\beq}{\begin{equation}}
\newcommand{\eeq}{\end{equation}}
\newcommand{\be}{\begin{equation}}
\newcommand{\ee}{\end{equation}}
\newcommand{\bea}{\begin{eqnarray}}
\newcommand{\eea}{\end{eqnarray}}
\newcommand{\ba}{\begin{array}{ccc}}
\newcommand{\ea}{\end{array}}

\newcommand{\bra}[1]{\left ( #1 \right)}
\newcommand{\sbra}[1]{\left [ #1 \right]}
\newcommand{\sqbra}[1]{\left | #1 \right|}
\newcommand{\cbra}[1]{\left \{ #1 \right\}}

\newcommand{\vphi}{\vec{\phi}}

\def\frac#1#2{{\textstyle{#1 \over #2}}}
\def\half{\frac{1}{2}}
\def\blangle{{\big\langle}}
\def\brangle{{\big\rangle}}

\def\Bx{{\mib x}}
\def\Bq{{\mib q}}
\def\vphis{\vphi^{\,2}}
\def\ns{^{\vphantom{*}}}
\def\ij{{\langle ij\rangle}}
\def\sss#1{{\scriptscriptstyle #1}}
\def\ssr#1{{\sss{\rm #1}}}
\def\CZ{{\mathcal Z}}
\def\CG{{\mathcal G}}
\def\sigQ{\sigma_{\ssr Q}}
\def\sigB{\sigma_{\ssr B}}
\def\sigV{\sigma_{\ssr V}}
\def\bvph{\vphantom{\int\limits_A^B}}

\begin{document}

\title{Dynamics and Conductivity Near Quantum Criticality}

\author{Snir Gazit}
\author{Daniel Podolsky}
\author{Assa Auerbach}
\affiliation{Physics Department, Technion, 32000 Haifa, Israel}
\author{Daniel P. Arovas}
\affiliation{Department of Physics, University of California at San Diego, La Jolla, California 92093, USA}

\date{\today }

\begin{abstract}
Relativistic O($N$)  field theories are studied near the quantum critical point in two space dimensions. We compute dynamical correlations
by large scale Monte Carlo simulations and numerical analytic continuation.  In the ordered side, the scalar spectral function
exhibits  a universal  peak at the Higgs mass. For $N=3$ and $4$ we confirm its $\omega^3$ rise at low frequency. On the disordered side,
the spectral function exhibits a sharp gap.  For $N$=2, the dynamical conductivity rises above a threshold at the Higgs mass (density gap),
in the superfluid  (Mott insulator)  phase.  For charged bosons, (Josephson arrays)  the  power law rise above the Higgs mass, increases
from two to four.  Approximate charge-vortex duality is reflected in the ratio of imaginary conductivities on either side of the
transition.  We determine the critical conductivity to be $\sigma^*_{\rm c} = 0.3 (\pm 0.1) \times 4e^2/h$. In the appendices, we describe
a generalization of the worm algorithm to $N>2$, and also a singular value decomposition error analysis for the numerical analytic
continuation.
\end{abstract}

\maketitle

\section{Introduction}

Relativistic O$(N)$ models describe the low temperature properties of diverse condensed matter systems: {\it e.g.,\/} quantum antiferromagnets, charge density waves, Josephson junction arrays, granular superconductors, and Bose condensates in optical lattices
\cite{fisher_boson_1989,CHN}. Many of these systems exhibit a quantum phase transition, as a function of a quantum tuning parameter,
between phases where the O($N$) symmetry is present and where it is spontaneously broken.

Far from criticality, the collective excitations in these systems are well understood. In the symmetric phase, there are $N$ massive modes.
In the broken symmetry phase, there are $N-1$ massless Goldstone modes and one massive amplitude (Higgs) mode\cite{Huber_Amplitude}.
The Higgs is actually a resonance, since it can decay into pairs of Goldstone modes. The shape of the resonance in two spatial dimensions
has recently attracted significant theoretical and experimental interest.  Weak coupling and $N=\infty$ diagrammatic expansions
\cite{Podolsky_visibility,Lindner_BadMetal} have shown that a careful choice of the correlation function ensures that the Higgs mode can be detected, without spurious contamination from massless Goldstone modes.  Indeed, the Higgs resonance has been directly observed in recent experiments of the Mott insulator to superfluid transition in optical lattices\cite{endres}.

Near the quantum critical point (QCP), the dynamics in two spatial dimensions are determined by a strongly coupled fixed point, thus 
precluding a simple description in terms of weakly interacting quasiparticles.  In the ordered phase, Goldstone's theorem still ensures the
existence of Goldstone modes, which are long lived at low energy.  By contrast, there is no corresponding protection for the Higgs mode.
A $1/N$ expansion of the scalar susceptibility \cite{dpss} and numerical quantum Monte Carlo (QMC) simulations \cite{pollet} provided the
first evidence for the survival of the Higgs near criticality.  Although the low frequency  spectral function near criticality is predicted
to be universal \cite{dpss,QPTSubir}, its determination requires  numerical computation.  Recently, this was undertaken by
large scale QMC simulations of the scalar susceptibility for the O(2) and O(3) models\cite{gazit_fate} and for the Bose Hubbard model
\cite{kun_BH}. The Higgs peak in the ordered phase was clearly identified.

In this paper we further study the dynamical properties of relativistic O($N$) models close to the quantum critical point at low temperature,
frequency, and zero wave vector.  We review in detail our previous work in which we computed the universal line shape of the scalar
susceptibility \cite{gazit_fate} and present new numerical results for O$(N)$ models with $N=2$, $3$, and 4. In particular, we perform a
careful analysis of the  low frequency behavior of the line shape in the ordered phase, where we confirm the $\omega^3$ rise for $N=3$ and $N=4$ predicted in
Ref. \onlinecite{Podolsky_visibility}.  For $N=2$ we cannot resolve the low frequency power law.  The scalar response
in the disordered phase exhibits a sharp threshold above a gap.

We present QMC and analytic results for the dynamical conductivity of the O$(2)$ model on both sides of the transition.
In the superfluid phase, we find a threshold-like behavior in the conductivity, which rises quadratically with frequency above the Higgs mass $m\ns_H$. In the insulator there is a low-frequency threshold in the conductivity appearing at twice the single particle gap $\Delta$, and a negative (capacitive) linear
dependence of the imaginary conductivity.  

Throughout the analysis we identify a number of universal constants that characterize the critical point.  These include ratios of quantities measured on mirror points on the ordered/disordered sides of the transition, such as
$m\ns_H/\Delta$ and $\Upsilon/\Delta$, where  $\Upsilon$ is the helicity modulus in the ordered phase (superfluid stiffness in the superfluid phase).
For $N=2$, we compute the high
frequency universal conductivity $\sigma_{\rm c}^*(\omega\gg T)$ in the quantum critical regime. In addition, we compute the universal
ratio $C/L$ between the low frequency capacitance in the insulator and the inductance in the superfluid to one loop order.  These results are consistent with  vortex-charge 
duality~\cite{fisher_duality} relations between the two sides of the QCP.

Our results are relevant to recent experiments which probe critical dynamics.  In cold atomic gases,
the Higgs mode has been excited by modulating the lattice potential near the superfluid to Mott transition\cite{endres}. 
Fast real time pump-probe response was used to see amplitude oscillations in charge density wave (CDW) systems\cite{Demsar_CDW,yusupov_CDW}.
Raman and neutron scattering have long identified a {\em ``two magnon peak''} in antiferromagnets\cite{Lyons_Raman,Parkinson_Raman,Fleury_2Mag,Elliott_2Mag,Shastry_Raman}. 
Within our theory, this peak is a Higgs mode which would soften at criticality. The conductivity in cold atom systems may be measured by
lattice phase modulations\cite{Giamarchi_Phase}. For Josephson junction arrays and granular superconducting films, Coulomb interactions
must be considered, as they give rise to massless two-dimensional plasmons.  We show that this increases the power law rise of the
conductivity above the Higgs threshold.  While our theory is for translationally invariant systems, some of the finite frequency zero
wave vector results may be a good starting point for understanding very recent results on disordered granular superconducting
films\cite{ShermanTHz}.

This paper is organized as follows. Section \ref{Sec:FT} presents the O$(N)$ field theory and the observables we study, together
with their expected scaling near the quantum critical point.  Section \ref{MoMe} introduces the discretized lattice model.
In Section \ref{secCES}, we locate the critical point as a function of cutoff parameters and compute the relevant energy scales near the
critical point.  In Section \ref{secSS}, we present the universal scaling functions of the scalar susceptibility. In Section \ref{secDC},
we compute the dynamical conductivity on both sides of the superfluid-Mott transition and present an approximate duality for the optical
conductivity.  Appendix \ref{AppA} describes the QMC algorithm in detail.  Appendix \ref{AppB} discusses the numerical analytical
continuation procedure and provides an error analysis of the kernel pseudo-inversion.  Finally, Appendix \ref{AppC} describes a
weak coupling analytic calculation of the conductivity.

\section{Field Theory and Scaling}\label{Sec:FT}

We will study microscopic systems with O$(N)$ symmetry whose long wave length and low energy universal properties near the QCP are captured
by a quartic field theory with relativistic dynamics\cite{QPTSubir}.  The field theory in 2+1 dimensional Euclidean space-time is given by
\begin{equation}
\begin{split}
\CZ &=\int\! \mathcal{D} \vphi \> e^{-S\left[\vphi\right]} \\
S\big[\vphi\,\big]&=\int\limits_\Lambda \!\! d^2\! x \, d\tau  \sbra{ \half\big(\pmu \vphi\big)^2 + \half \mu \vphi^{\,2} +
g \big(\vphi^{\,2}\big)^{\!2}}
\label{eq:ContModel}
\end{split}
\end{equation}
The fields $\vec{\phi}$ are $N$-component real fields, $\pmu=\{\partial_\tau,\partial_x,\partial_y\}$, and $\Lambda$ is an ultraviolet
cutoff.   Examples of physical realizations include the superfluid to Mott insulator transition of lattice bosons at commensurate fillings
\cite{fisher_boson_1989} for $N=2$ and the N\'{e}el to singlet transition of dimerized Heisenberg antiferromagnets\cite{CHN} for $N=3$.

The system undergoes a quantum phase transition as the quantum tuning parameter $g$ is varied. For $g<g\ns_{\rm c}$ the O$(N)$ symmetry
in spontaneously broken as the field obtains a non zero expectation value $\blangle\vphi\,\brangle\ne 0$. The ordered phase is then
characterized by $N-1$ massless Goldstone modes and a single gapped Higgs mode. For $g>g\ns_{\rm c}$ the system is disordered and contains
$N$ massive modes with excitation gap $\Delta(g)$. A dimensionless QCP tuning parameter is defined by
$\delta g = (g-g\ns_{\rm c})/g\ns_{\rm c}$.

We study two dynamical observables: the scalar susceptibility and the dynamical conductivity. For completeness we define these observables and discuss their expected scaling behavior and experimental realizations.

\subsection{Scalar susceptibility}

The scalar susceptibility describes the response function of experimental probes that are sensitive to the amplitude of the order parameter,
but not to its direction \cite{Podolsky_visibility}. As an example, the scalar susceptibility has been recently measured in experiments on
cold atoms on optical lattices near the Mott insulator-superfluid transition\cite{endres}. The experimental protocol consists of modulating
the optical lattice depth at a fixed frequency and measuring the energy absorbed using an {\it in-situ\/} imaging technique.  Here, the
perturbation modulates the condensate density, which is proportional to the square of the order parameter amplitude.

The scalar susceptibility is defined as the correlation function of the order parameter amplitude squared:
\begin{equation}
\begin{split}
\chi_s(\tau) &= \int\! \!d^2\!x \> \Big(\blangle\vphis_{x,y,\tau}\,\vphis_{\bf{0}}\brangle-\blangle\vphis_{\bf{0}}\brangle^2 \Big) \\
\chi_s(i\omega_m) &=\int_0^\beta\!\! d \tau\> e^{i\omega_m\tau}\chi_s(\tau) 
\label{eq:scalarSuspImag}
\end{split}
\end{equation}
The real frequency spectral function is obtained by analytic continuation of Eq.~\eqref{eq:scalarSuspImag}
\beq
\chi''_s(\omega)=-\Imag \chi_s(i\omega_m\to \omega+i0^{+})
\label{eq:AnalyticCont}
\eeq

Scaling arguments indicate that the expected low energy form of Eq.~\eqref{eq:scalarSuspImag} near the QCP is\cite{dpss}:
\bea
\chi_s(\omega/\Delta)\sim {C}+{\mathcal{A}}_\pm \Delta^{3-2/\nu} \Phi_{\pm}(\omega/\Delta)
\label{eq:scalarSuspScal}
\eea 
Where $\Delta\sim |\delta g|^\nu$ is the gap in the disordered phase, $\nu$ is the correlation length critical exponent, and
$\Phi_-$ ($\Phi_+$) is a universal function of $\omega/\Delta$ on the ordered (disordered) side of the transition. The non-universal constant $C$
is real, and is a regular function of $g$ across the transition. The ordered phase is gapless due to the presence of
Goldstone modes.  In order to provide a well-defined energy scale that characterizes fluctuations on the ordered phase $ (\delta g <0)$,
we use the gap at the mirror point $-\delta g$ across the transition. 

\subsection{Conductivity}

The dynamical conductivity measures the response to an external gauge field. Our analysis will be restricted to the $N=2$ case, as is
relevant to dynamical conductivity measurements in  superconductors and also to neutral cold atoms probed by optical lattice phase
modulations  \cite{Tokuno_LatticeOptCond} . To simplify the analysis we write the two scalar fields in Eq.~\eqref{eq:ContModel} as a
single complex field $(\phi_1,\phi_2) =\sqrt{2}\,(\Real \Psi,\Imag \Psi)$. We introduce the gauge field $A_\mu$ through minimal coupling
$\pmu \Psi \to \left(\pmu+ie^* A_\mu\right)\Psi$ for a field $\Psi$ carrying charge $e^*$.

The current is obtained by differentiating the action with respect to $A_{\mu}$, {\it viz.\/}
\begin{equation}
\begin{split}
\avbra{J_\mu}&={\delta S(A)\over\delta A_\mu}\\
&= ie^* \blangle\Psi^*\pmu\Psi-\Psi\pmu\Psi^*\brangle + 2 e^{*2} A_\mu \blangle |\Psi|^2 \brangle\ ,
\end{split}
\end{equation}
from which we derive the response function:
\begin{align}
\Pi_{\mu\nu}(x,x')&=\frac{\delta}{\delta A_\nu(x')}\,\avbra{J_\mu(x)}\big|\ns_{A=0}\\
&= \avbra{J_\mu(x) \,J_\nu(x')}+2 e^{*2} \avbra{\sqbra{\Psi}^2}\,\delta_{\mu\nu}\,\delta(x-x')\ . \nonumber
\end{align}
The first term is the paramagnetic response kernel $\Pi^{\rm P}_{\mu\nu}(x,x')=\avbra{J_\mu(x)\, J_\nu(x')}$, and the second term is the
diamagnetic response.  The conductivity is then given by
\beq
\sigma(i \omega_m)=-{1\over\omega_m}\,\Pi_{xx}(i\omega_m,q=0)\ .
\eeq
As in Eq.~\eqref{eq:AnalyticCont}, the real frequency dynamics is obtained by analytic continuation,
\beq
\sigma(\omega)= \sigma(i \omega_m \to \omega + i \epsilon)\ .
\label{eq:AnalyticContCond}
\eeq

Remarkably, in 2+1 dimensions the scaling dimension of the conductivity is zero\cite{fisher_UniCond}. As a result, near the critical point the conductivity has the scaling form \cite{fisher_UniCond,damle_UniCond},
\beq
\sigma(\omega) =\sigQ\, \Sigma_{\pm} (\omega/\Delta)\ .
\label{eq:condscal}
\eeq
Here $\sigQ=e^{*2}/h$ is the quantum of conductance and $\Sigma_{\pm}$ are dimensionless universal functions of $\omega/\Delta$ for the
disordered (+), and ordered (-) phases. 

\section{Model and Methods}\label{MoMe}

In order to simulate the continuum field theory Eq.~\eqref{eq:ContModel} we consider the following discrete lattice model:
\begin{equation}
\begin{split}
\CZ &=\int\!\!{\cal D} \vphi \> e^{-S[\vphi\,]} \\
S&= \sum_\ij \vphi_i \cdot \vphi_{j} +\mu \sum_i  \big|\vphi_i\big|^2 + g \sum_i  \big|\vphi_i\big| ^4\ .
\label{eq:latModel}
\end{split}
\end{equation} 
Here $\vec{\phi}$ is an $N$ component scalar field, residing on the sites of cubic lattice of linear size $L$ with periodic boundary
conditions.  The model is the same as that considered in Ref. \onlinecite{gazit_fate}, as seen by rescaling $\vphi_i\to g^{-1/2}\vphi_i$.
The long wavelength properties of Eq.~\eqref{eq:latModel} are captured by the field theory Eq.~\eqref{eq:ContModel}.  This model can be
interpreted either as a quantum mechanical partition function in discrete $2+1$ Euclidean space-time dimensions, or as a classical
statistical mechanics model in three dimensions. Near the phase transition between ordered and disordered phases, this minimal model
captures the critical properties of Eq.~\eqref{eq:ContModel} while explicitly treating space and time on an equal footing and preserving
exact particle-hole symmetry ($\Psi\to \Psi^*$) for the $N=2$ case. 

Next we define the discrete lattice version of the continuum observables. The scalar susceptibility is given by
\bea
\chi_s(\tau) &=& \sum_{x,y} \avbra{\vphis_{x,y,\tau}\,\vphis_{\bf{0}}}-\avbra{\vphis_{\bf{0}}}^2 \ .
\label{eq:latSS}
\eea
To define the conductivity it is easier to consider the $U(1)$ symmetric complex field analog model of the $N=2$ scalar field,
\begin{align}
\CZ &=\int\! \mathcal{D}\Psi \, \mathcal{D}\Psi^*\> e^{-S\left[\Psi,\Psi^*\right]} \label{eq:latModel2}\\
S&=\sum_\ij \left(\Psi_i^* \Psi\ns_j+ \Psi\ns_i \Psi^*_j\right)
+2\mu \sum_i  |\Psi_i|^2  + 4g\sum_i |\Psi_i| ^4\ .\nonumber
\end{align}
We introduce the gauge field $A_\mu(i)$ through Peierls substitution $\Psi_i^*\Psi\ns_{i+\mu} \to
\Psi_i^* \Psi\ns_{i+\mu}\,e^{i e^* A_\mu(i)}$. The current is then
\begin{equation}
J_\mu(i) = {\delta S\over\delta A_\mu(i) }=ie^* \avbra{\Psi_i^*\Psi\ns_{i+\hat{\mu}}\,e^{i e^* A_\mu(i)}- {\rm c.c.}} 
\end{equation}
and the response function,
\begin{equation}
\begin{split}
\Pi_{\mu\nu}(i,j) &= \frac{\delta}{\delta A_\nu(j)}\, \avbra{J_\mu(i)}\Big|\ns_{A=0}\\
&= \Pi^{\rm P}_{\mu\nu}(i,j) +K\delta_{\mu\nu}\,\delta_{i,j}\ .
\end{split}
\end{equation}
$\Pi^{\rm P}_{\mu\nu}(i,j)=\avbra{J_\mu(i)\, J_\nu(j)}$ and $K=-e^{*2}\avbra{\Psi_i^*\Psi\ns_{i+\hat{\mu}}+{\rm c.c.}}$ are
the lattice versions of, respectively, the paramagnetic and the diamagnetic response.

The simplicity of our model allowed us to simulate large system sizes, up to $L=200$. Considering such large systems enabled us to
accurately track the critical properties near the QCP. This is especially important in the ordered phase where the system is gapless
and the dynamical response functions have power-law behavior. We implemented the highly efficient ``worm algorithm" 
\cite{prokofev_worm_2001}, sampling from a dual closed loops representation. The correlation time of the worm algorithm scales well
with system size, suppressing the critical slowing down near the transition. We also extend the work of Ref.
\onlinecite{prokofev_worm_2001} to treat general O$(N)$ models with $N>2$.  Details of the QMC algorithm can be found in appendix A.    
We compared our numerical results against previous QMC
studies of O$(N)$ models\cite{hasenbusch_eliminating_2001,hasenbusch_high-precision_1999} and with analytically solved limits and
found good agreement within error bars. 

A key ingredient of our analysis is the numerical analytic continuation of imaginary time QMC data to real frequency spectral functions.
To do so we have to invert the relation,
\begin{equation}
\CG (i\omega_m)= \int\limits_{0}^{\infty} \!{d \nu\over\pi} \> {2\nu\over\omega_m^2+\nu^2}\>A(\nu)\ .
\end{equation}
Here $\CG (i\omega_m)$ is a correlation function in Matsubara frequency space, evaluated by the QMC simulation, and $A(\nu)$ is the
spectral function. However, the kernel has very small singular value eigenvalues, and the inversion can unwittingly amplify the statistical QMC noise in $\CG (i\omega_m)$.  
A detailed discussion of methods which can circumvent  these artifacts is presented in Appendix B.

\begin{figure}[t!]
\includegraphics[width=0.475\textwidth]{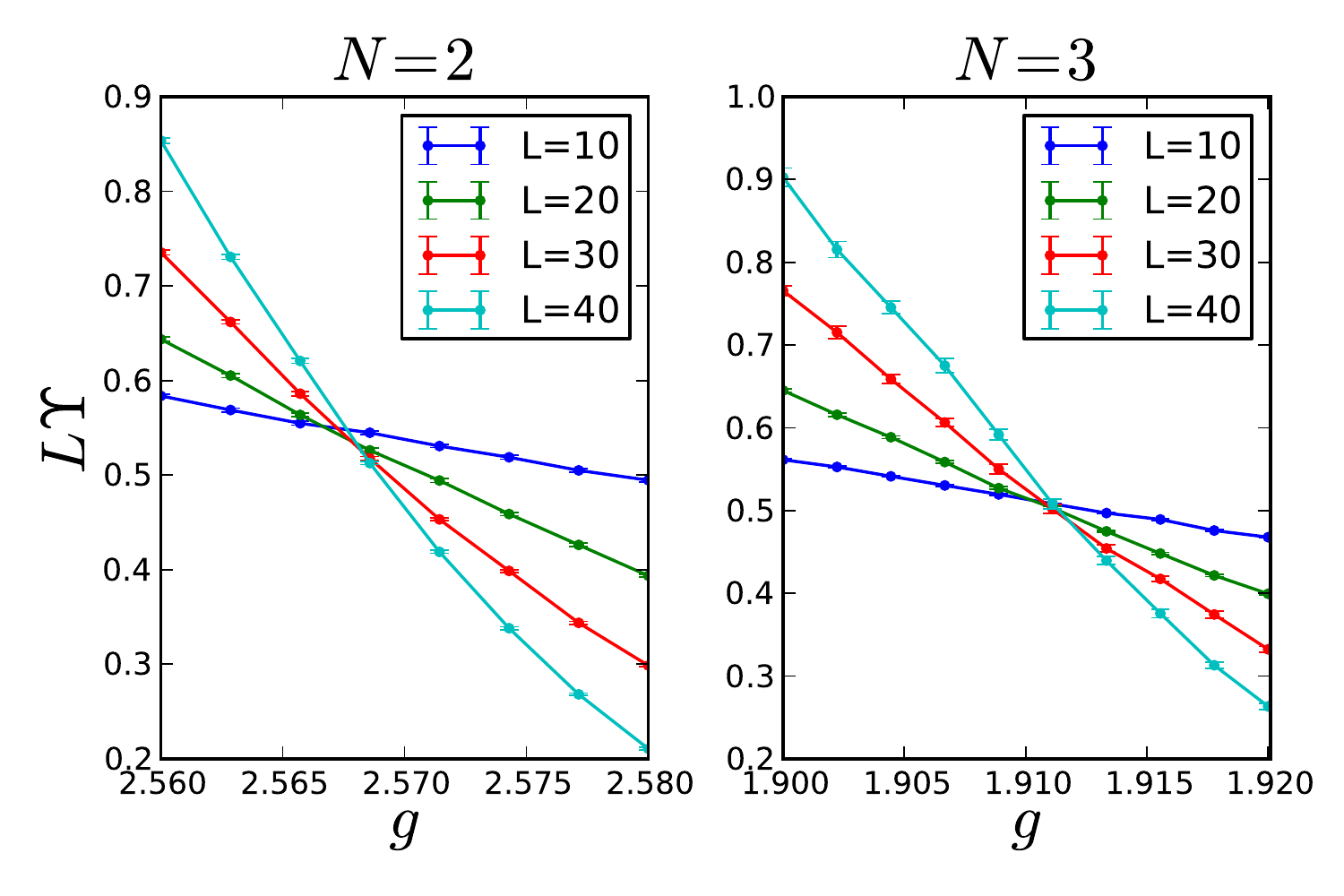}
\caption{Curves of $L\Upsilon$ for an sequence of increasing system size $L$ for O$(N=2,3)$ models. The curve cross at a single point,
from which we determine the value of $g_{\rm c}$. Here we take $\mu=-0.5$ and $g_{\rm c}=2.568(2)$ for the $N=2$ case and
$g_{\rm c}=1.912(2)$ for $N=3$.}
\label{fig:findgc}
\end{figure}

\section{Critical Energy Scales}\label{secCES}

\subsection{Determination of the critical coupling}

In order to study critical properties it is necessary to locate the QCP with high accuracy.  We determine the critical coupling by finite
size scaling analysis of the helicity modulus of the 2+1-dimensional quantum model.  The helicity modulus $\Upsilon$  is defined by
$\Upsilon\equiv\frac{1}{L}\frac{\partial^2 \ln {\CZ}(\varphi)}{\partial\varphi^2}|_{\varphi=0}$ where $\CZ(\varphi)$ is the partition
function in the presence of a uniform phase twist $\varphi$.  Near the critical point, $\Upsilon L $ is a universal constant, with only
next-to-leading order corrections in the system size $L$.\cite{fisher_UniCond,Sorensen_UniCond}  The critical coupling is then determined
from the crossing point of $L \Upsilon$ for a sequence of increasing system sizes $L$. Illustrative examples for $N=2$ and $N=3$ are shown
in Fig. \ref{fig:findgc}. Curves for different system sizes cross at a single point with little variation with system size, allowing us
to determine the critical coupling accurately.

We studied a few different parameter sets $(g_{\rm c},\mu_{\rm c})$  which are shown in Table~\ref{tab:params}.  The use of multiple sets of model parameters for $N=2$
allowed us to test the universality of our results. In most cases we tuned the transition by varying $g$, except in the case of dynamical
conductivity, where we varied $\mu$.

\begin{table}[t!]
\begin{tabular}{| l | l | l | l | }
  \hline                        
  Model & $N\quad$  & Model parameters$\quad$ & Critical coupling $\quad$\\
  \hline
  A &  2 & $\mu=-0.5$ & $g_{\rm c}=2.568(2)$ \\
  \hline
  B &  2 & $\mu=-2$ & $g_{\rm c}=3.908(2)$ \\
  \hline  
  C &  2 & $g=7.6923$ & $\mu_{\rm c}=-5.883(2)$ \\
  \hline  
  D &  3 &  $\mu=-0.5$ & $g_{\rm c}=1.912(2)$ \\
  \hline  
  E &  4 &  $\mu=-0.5$ & $g_{\rm c}=1.516(2)$ \\
  \hline  
\end{tabular}
\caption{List of model parameters studied, along with their critical couplings.}
\label{tab:params}
\end{table}

\subsection{Excitation gap in the disordered phase}

\begin{figure}[t!]
\includegraphics[width=0.475\textwidth]{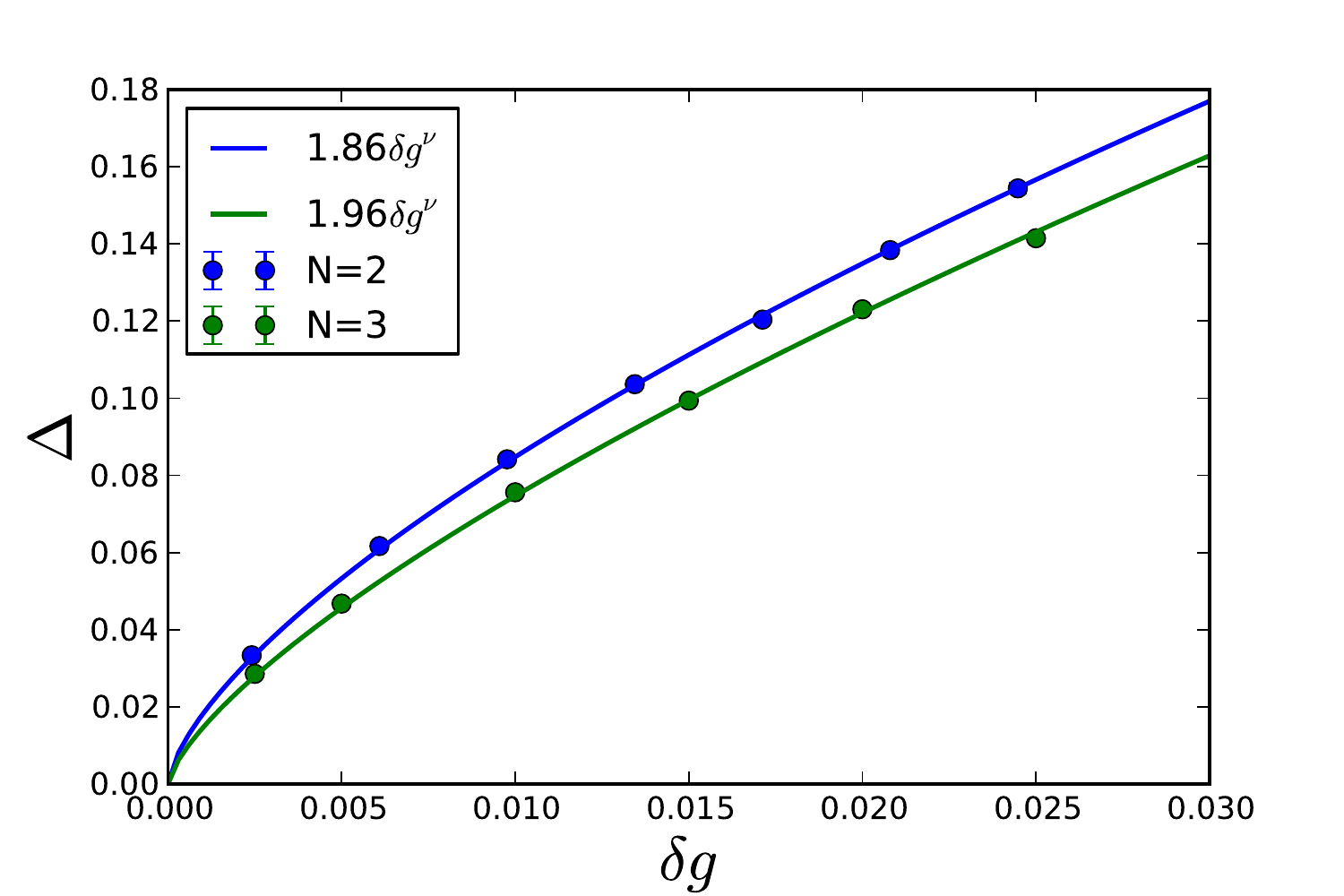}
\caption{Scaling of the gap $\Delta(\delta g)$ in the disordered phase for  $N=2,3$ and $\mu=-0.5$.  Fitting to the scaling form
$\Delta= \Delta_0 (\delta g)^\nu$ gives $\Delta_0$=1.86(1) for $N=2,\mu=-0.5$ and $\Delta_0$=1.96(1) for $N=3, \mu=-0.5$. Error bars
are smaller than the symbols.}
\label{fig:deltafit}
\end{figure}

The gap in the disordered phase provides a reference energy scale for all dynamical properties.  It can be extracted with high precision
from the zero momentum two point Green's function\cite{Gap2DBose},
\beq
G(\tau)=\sum_{x,y}\avbra{\vphi_{x,y,\tau}\cdot\vphi_{\bf{0}}}\ ,
\eeq
without recourse to analytic continuation.  At large imaginary times, $G(\tau)$ is expected to behave as
\beq
 G(\tau) \sim e^{-\Delta \tau} + e^{-\Delta(\beta-\tau)}\ .
\eeq
The gap $\Delta$ is evaluated by a fit to the above functional form. The evolution of the gap near the QCP is depicted in
Fig. \ref{fig:deltafit} for $N=2,3$. The gap softens as $\delta g\to 0$ according to the scaling form
$\Delta (g) \sim \Delta_0 \,(\delta g)^\nu$, from which we extract $\Delta_0$.  For the correlation length exponent $\nu,$ we use
values determined in previous high accuracy simulations\cite{hasenbusch_high-precision_1999,hasenbusch_eliminating_2001}:  
$\nu_2=0.6723(3)$, $\nu_3=0.710(2)$, and $\nu_4=0.749(2)$ for $N=2$, $N=3$, and $N=4$ respectively. 

We validated our results by performing a similar analysis of the long imaginary time form of the scalar susceptibility \cite{gazit_fate}
$\chi_s(\tau) \sim \tau^{-1}\,e^{-2 \tau\Delta}$.  We found good agreement between the two approaches. 

\subsection{Helicity modulus in the ordered phase}
\label{sec:stiffness}

In two spatial dimensions, the helicity modulus is an energy scale that can be used to characterize the ordered phase.  For $N=2$ ($N=3$) it plays the role of the superfluid stiffness (spin stiffness).  Similarly to the gap in the disordered phase, the helicity modulus  near the QCP vanishes according to the scaling behavior $\Upsilon=\Upsilon_0 (\delta g)^\nu$. The ratio $\Upsilon_0/\Delta_0$ is universal. We find $\Upsilon_0/\Delta_0=0.44(1)$ for $N=2$ and $\Upsilon_0/\Delta_0 =0.34(1)$ for $N=3$.

\section{Scalar Susceptibility}
\label{secSS}

In the following,  the universal scaling functions of the scalar susceptibility are computed for both phases.  

\subsection{Matsubara frequency universal scaling function }

\begin{figure}[b!]
\includegraphics[width=0.475\textwidth]{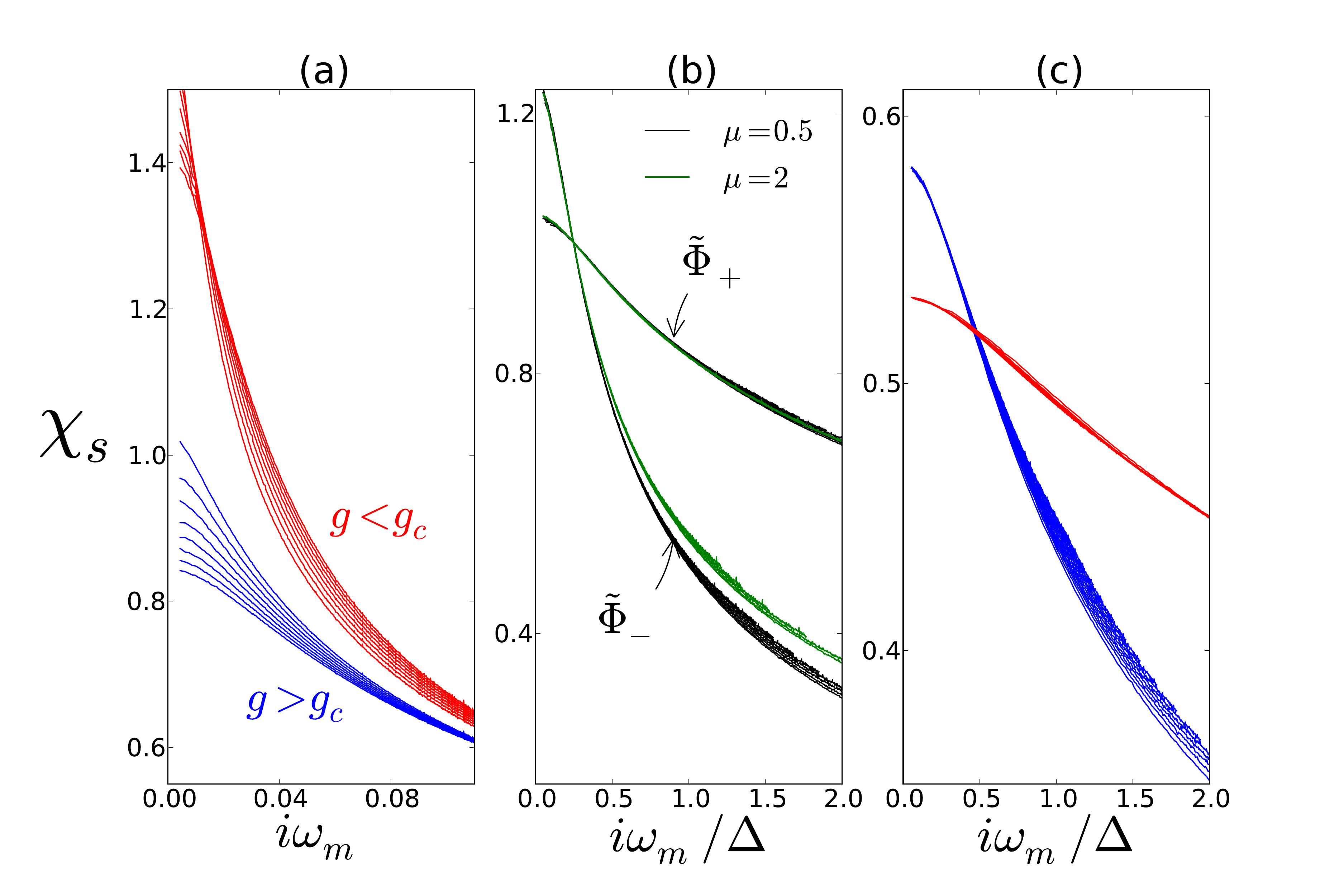}
\caption{{\bf (a)} The scalar susceptibility $\chi_s(i \omega_m)$ for $N=2$.  The curves correspond to different values of $\delta g$ below
and above the phase transition. {\bf (b,c)} universal scaling function after rescaling for $N=2,3$. In (b) we show the scaling function for
two crossing points of the phase transition. The two rescaled curves agree very well, especially at low frequencies. Simulations were performed
with $\mu=0.5$ and $\mu=2$ for $N=2$ and  $\mu=0.5$ for $N=3$}
\label{fig:imagecollapse}
\end{figure}

In Fig. \ref{fig:imagecollapse}(a) numerical results for the  $N=2$ scalar susceptibility $\chi_s(i\omega_m)$ as a function of Matsubara frequency are presented for
both phases.  The scaling form Eq.~\eqref{eq:scalarSuspScal} applies also to the correlation
function in Matsubara space. The universal scaling function $\Phi(i\omega_m)$ is then computed by rescaling the $\chi_s(i\omega_m)$ curves
according to Eq.~\eqref{eq:scalarSuspScal}.  The collapse requires the extraction of the non-universal real constant $C$, which is expected
to be a smooth function of $\delta g$.  We find $C$ by fitting $\chi_s(i\omega_m)$ at small $\omega_m$ to a polynomial in $\delta g$, and
then subtracting it from $\chi_s(i\omega_m)$.  The $\omega$ axis is then rescaled by $\Delta$ and the vertical axis is rescaled by
$\Delta^{3-2/\nu}$.  

\begin{figure}[t!]
\includegraphics[width=0.5\textwidth]{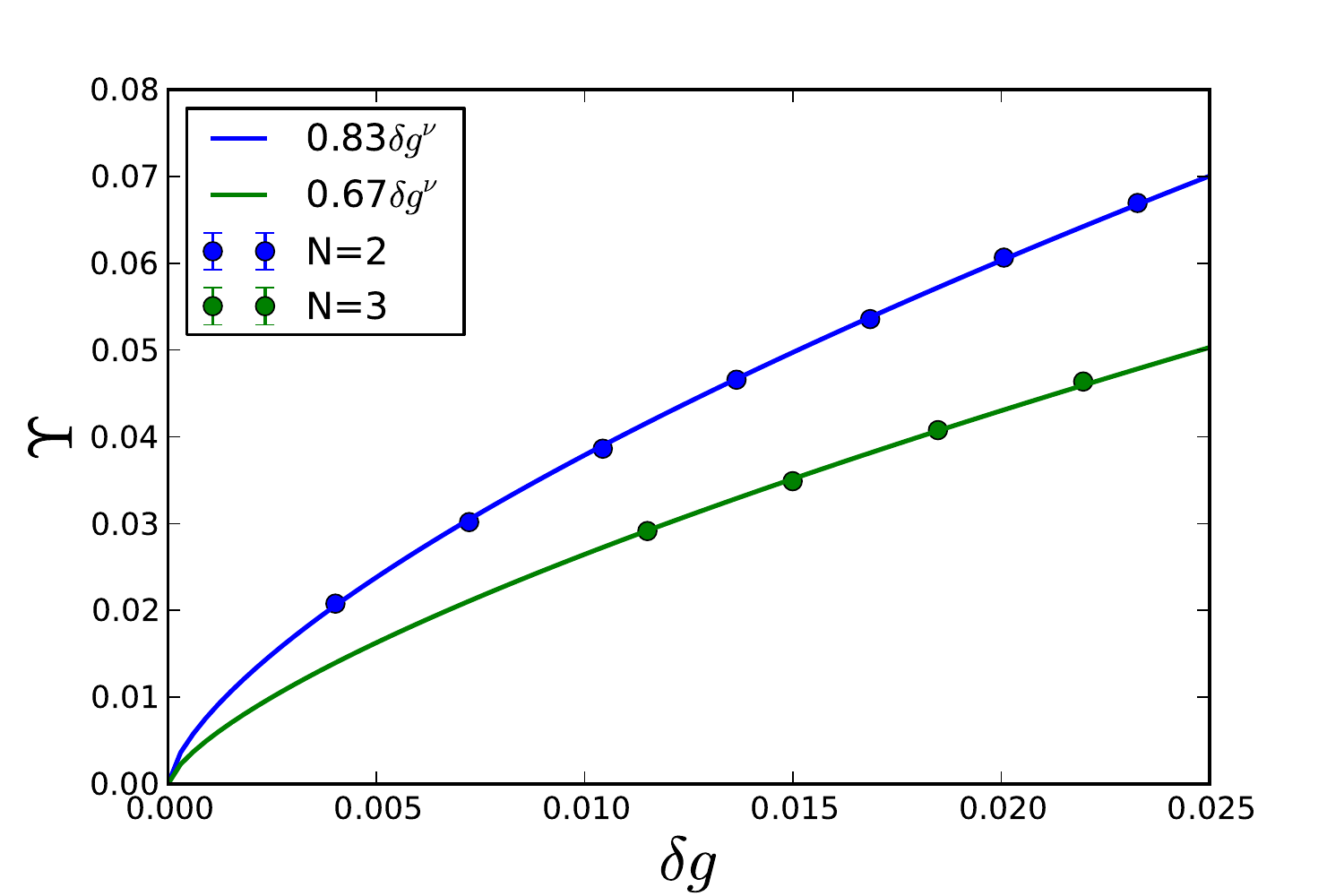}
\caption{Scaling of the helicity modulus $\Upsilon(\delta g)$ in the ordered phase for  $N=2,3$ and $\mu=-0.5$.  Fitting to the scaling
form $\Upsilon= \Upsilon_0 (\delta g)^\nu$ gives $\Upsilon_0$=0.83(1) for $N=2,\mu=-0.5$ and $\Upsilon_0$=0.67(1) for $N=3, \mu=-0.5$.
Error bars are smaller than the symbols.}
\label{fig:helicityfit}
\end{figure}

Figure \ref{fig:imagecollapse}(b,c) shows  the scaling procedure for $N=2,3$. The curves collapse into two universal
functions $\Phi_\pm(i\omega_m)$. To test the universality of our results we repeated the scaling analysis at a different
crossing point of the phase transition for the $N=2$ case. The results are presented in Fig.~\ref{fig:imagecollapse}(b). The scaled curves
for both sets of critical couplings agree very well, especially for low frequencies. This provides a stringent test for the consistency of our analysis.

\subsection{Real frequency universal scaling function }

Next we examine the imaginary part of the retarded response function $\chi_s^{\prime\prime}(\omega)$ obtained from analytic continuation of
$\chi_s(i\omega_m)$. To extract the universal part of the line shape we rescale the $\omega$ axis by $\Delta$ and the vertical axis by
$\Delta^{3-2/\nu}$. Note that this rescaling is done without any free fitting parameters, since the real constant $C$ in Eq.~\eqref{eq:scalarSuspScal} drops out from the spectral function. 

\begin{figure}[t!]
\includegraphics[width=0.475\textwidth]{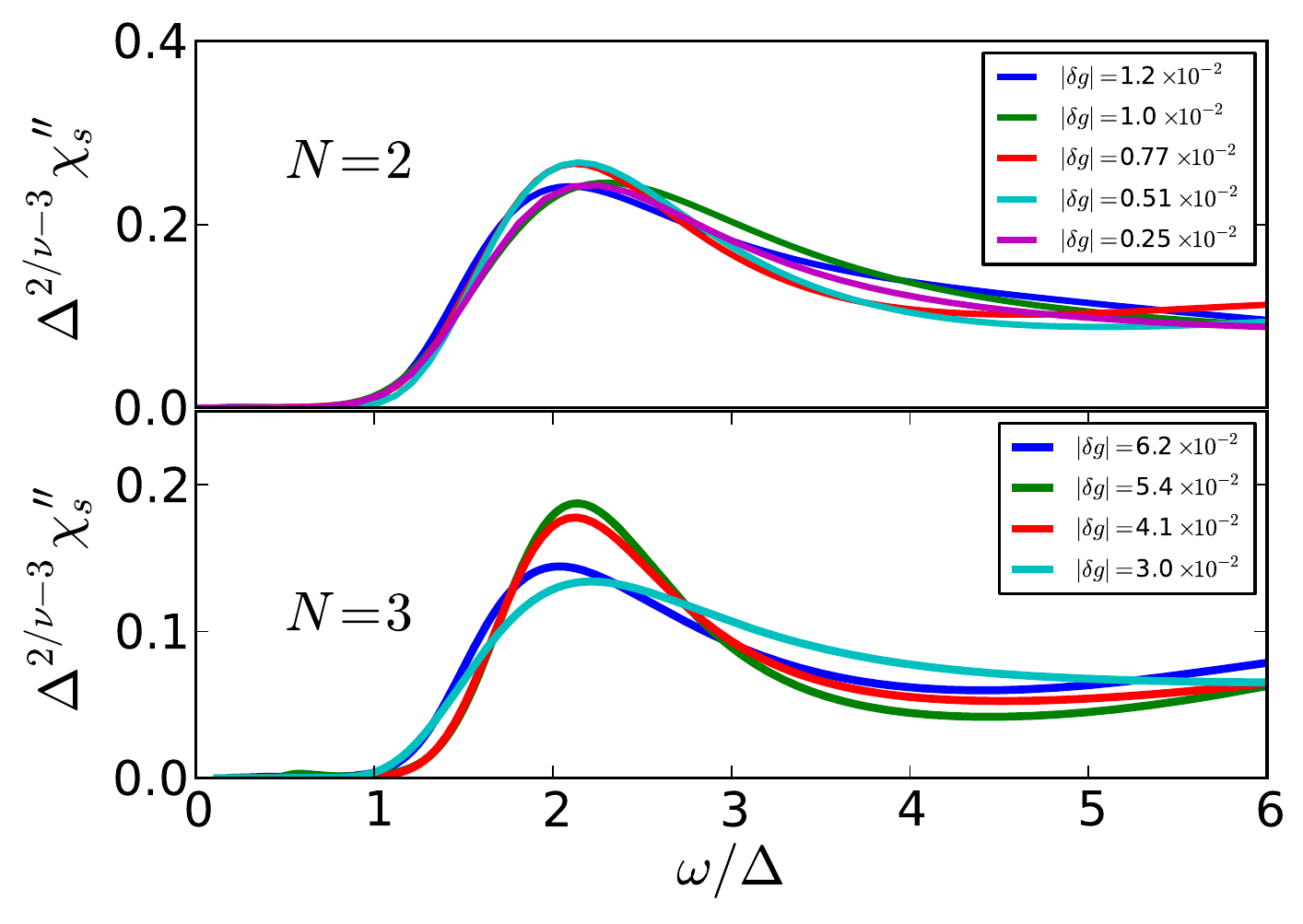}
\caption{$\chi_s''(\omega)$ in the ordered phase for $N=2$ and 3. We scale the curves according to Eq. \eqref{eq:scalarSuspScal} for a
range of tuning parameters $\delta g$ near the critical point.}
\label{fig:realcollapseOrd}
\end{figure}

The rescaled line shape in the ordered phase is shown in Fig.~\ref{fig:realcollapseOrd} for $N=2$ and $N=3$. Curves for different values of $\delta g$ collapse into a single universal line shape especially at low frequencies. The line shape contains a clear peak that can be associated with the Higgs mode.  Our analysis demonstrates that the Higgs peak is a universal feature in the spectral function that survives as a resonance arbitrarily close to the critical point. 

Some universal values can be obtained by  this analysis. For example, we consider the ratio between the Higgs mass in the ordered phase, defined by the maximum in $\chi_s''(\omega)$, and the gap in the
disordered phase at mirror points across the transition.  This ratio is found to be $m\ns_H/\Delta=2.1(3)$ and  $m\ns_H/\Delta=2.2(3)$ for $N=2$
and $N=3$ respectively.  We also obtain the fidelity $F=m\ns_H/\Gamma$, where $\Gamma$ is the full width at half-maximum. We measure
$\Gamma$ with respect to the leading edge at low frequency, since at low frequencies there is less contamination from the high frequency
non-universal spectral weight. Since the entire functional form of the line shape is universal, $F$ is a universal constant that
characterizes the shape of the peak. We find $F=2.4(10)$ for $N=2$ and  $F=2.2(10)$ for $N=3$.  

The rescaled spectral function in Fig.~\ref{fig:realcollapseOrd} shows higher variability at high frequencies than at low frequencies.
We attribute this to contamination from the non universal part of the spectrum and to systematic errors introduced by the maximum
entropy (``MaxEnt") regularization of the analytic continuation, which is  noisy in this regime.

\begin{figure}[t!]
\includegraphics[width=0.475\textwidth]{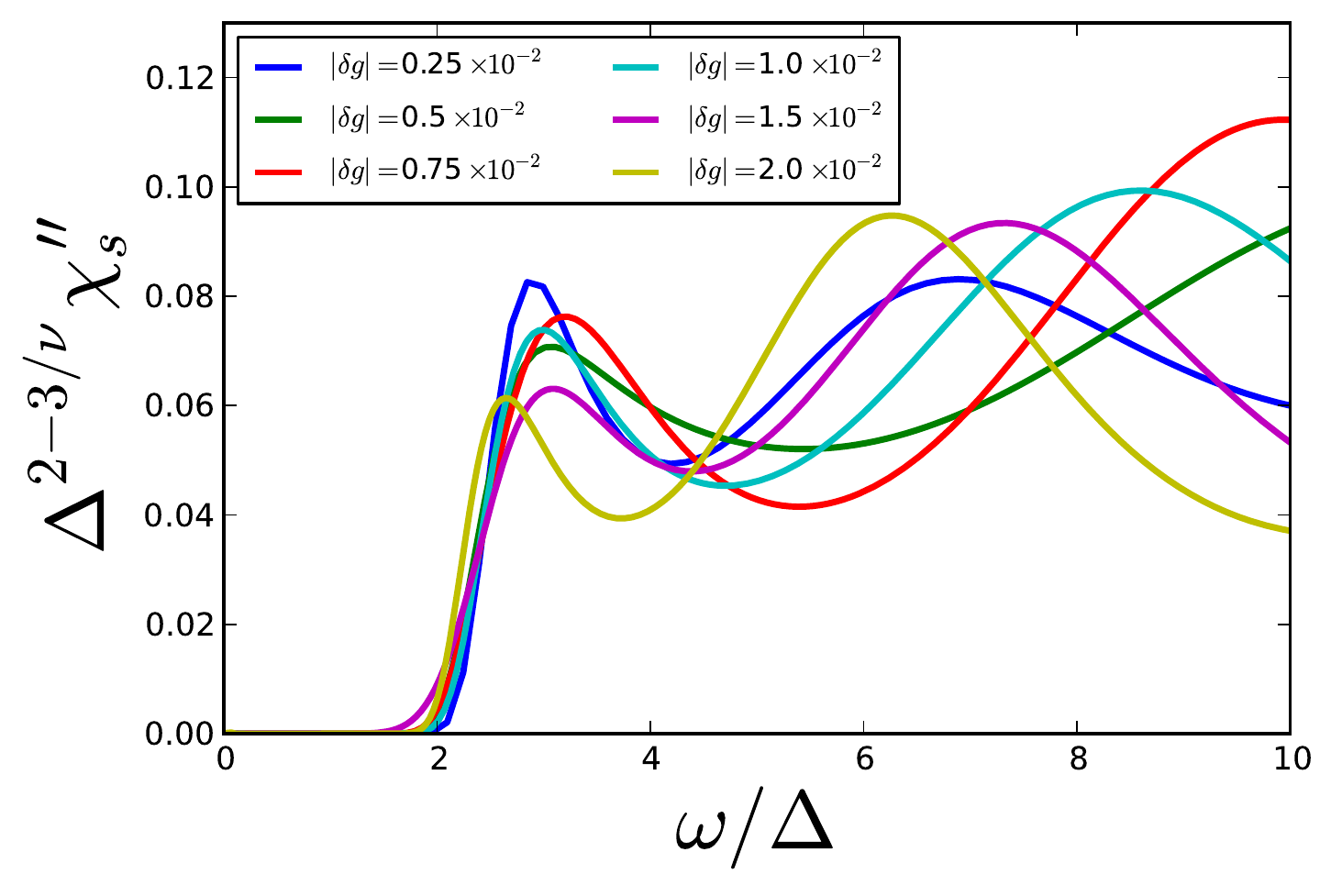}
\caption{$\chi_s''(\omega)$ in the ordered phase for $N=2$. We scale the curves according to Eq. \eqref{eq:scalarSuspScal} for a range
of tuning parameters $\delta g$ near the critical point.}
\label{fig:realcollapseDis}
\end{figure}

\begin{figure}[b!]
\includegraphics[width=0.475\textwidth]{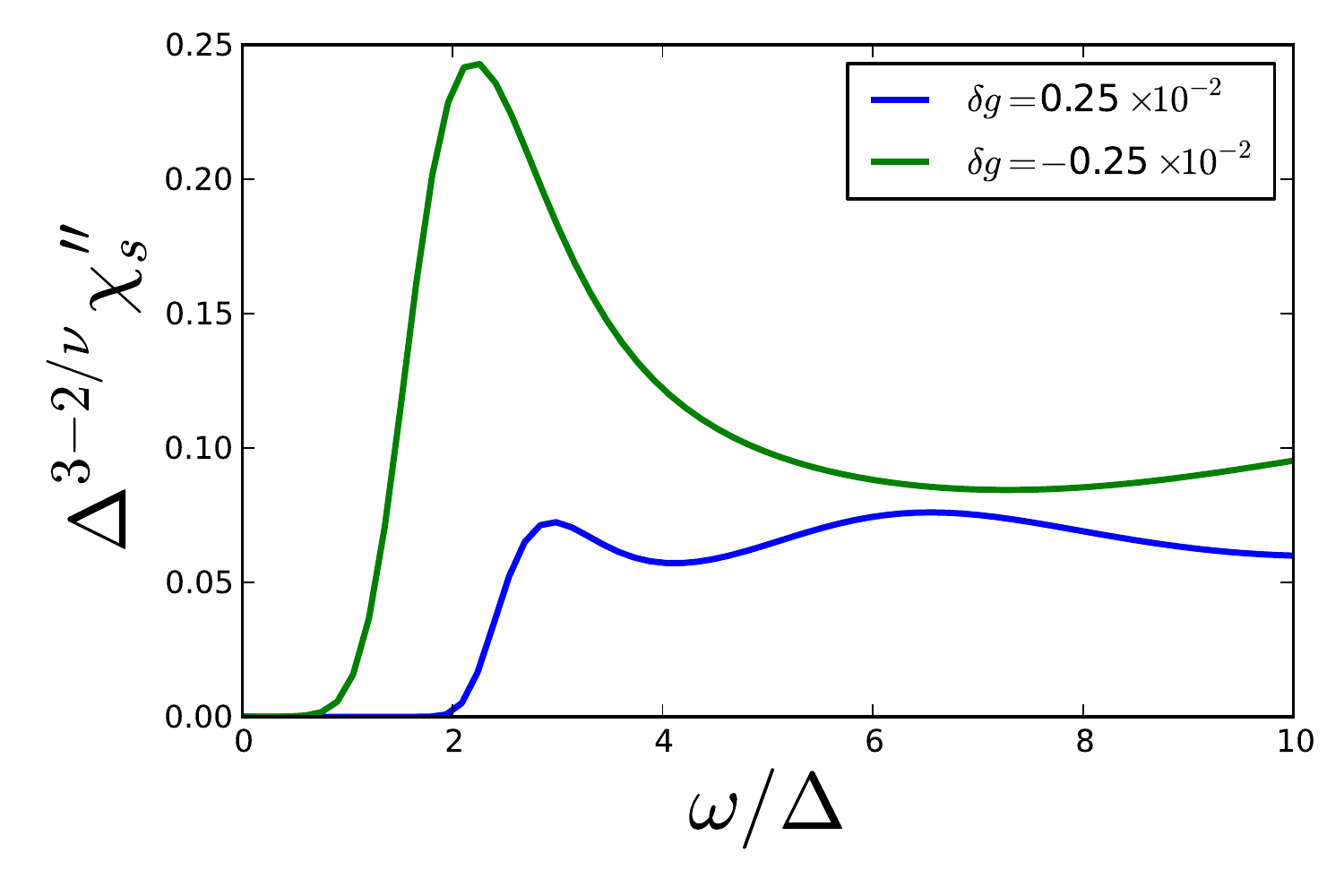}
\caption{Comparison of the scalar susceptibility line shape, $\chi^{\prime\prime}_s(\omega)$, on mirror points across the phase transition 
or $N=2$. The blue green curve corresponds to disordered phase and the green curve to the ordered phase.}
\label{fig:realcollapseComp}
\end{figure}

In Fig.~\ref{fig:realcollapseDis} we plot the rescaled line shape in the disordered phase for $N=2$. The universal spectral function is
gapped for $\omega<2\Delta$ and rises sharply above the threshold. This behavior is in accordance with analytic predictions\cite{dpss}
and with previous QMC numerical simulation \cite{Kun}.  Previous studies found a Higgs-like resonance in the disordered phase above the
threshold \cite{pollet,Kun}.  However, we find that the peak seen in Fig.~\ref{fig:realcollapseDis} at $\omega/\Delta\approx 3$ is very
shallow relative to the background spectral weight.  Thus we do not consider this to be conclusive evidence of a resonance.  We note that 
numerical analytic continuation  tends to produce oscillatory behavior near sharp features of the spectral function\cite{beach_maxEnt} and hence it is possible that the 
shallow peak might be an artifact of such an effect.
For comparison, in Fig. \ref{fig:realcollapseComp} we show representative curves for the line shape on mirror points of the transition.
If a resonance is at all present in the disordered phase, it is much less pronounced than in the ordered phase.

\subsubsection{Asymptotic power law decay of the scalar susceptibility}

In the ordered phase, the low frequency rise of the scalar susceptibility was predicted~\cite{ssrelax,Podolsky_visibility,dpss} to be
\begin{equation}
\Phi''_-(\omega) \sim \left(\omega/\Delta\right)^3 \quad,\quad  \omega\ll \Delta \ll 1.
\label{eq:UniversalOrd}
\end{equation}
The $\omega^3$ rise  is due to the decay of a Higgs mode into a pair of Goldstone modes.  Equation (\ref{eq:UniversalOrd}) transforms into
the large imaginary time asymptotic form  $\chi_s\bra{\tau} \sim 1/\tau^{4}$. Hence, to test Eq.~\eqref{eq:UniversalOrd} we examine the
large $\tau$ behavior of $\chi_s(\tau)$. We note that this approach does not rely on analytic continuation, enabling us to study the low
frequency dynamics in a numerically stable and well controlled manner.

\begin{figure}[t!]
\includegraphics[width=0.475\textwidth]{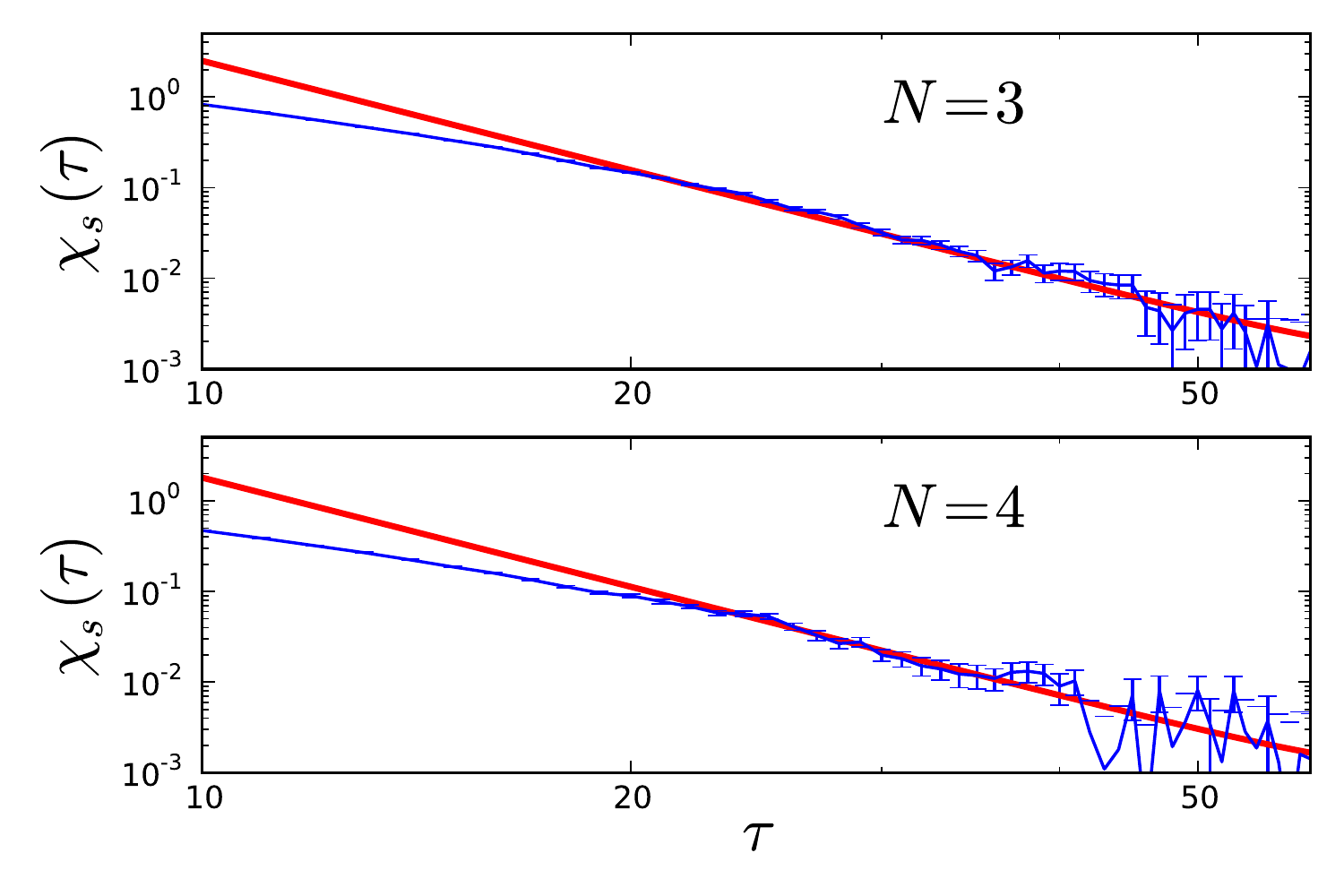}
\caption{Log-log scale plot for $\chi_s(\tau)$ in the ordered phase. For $N=3,4$  we indeed find the asymptotic behavior $\chi_s(\tau)\sim 1/\tau^{4}$ to agree within the error bars. }
\label{fig:tau4}
\end{figure}

\begin{figure}[t!]
\includegraphics[width=0.475\textwidth]{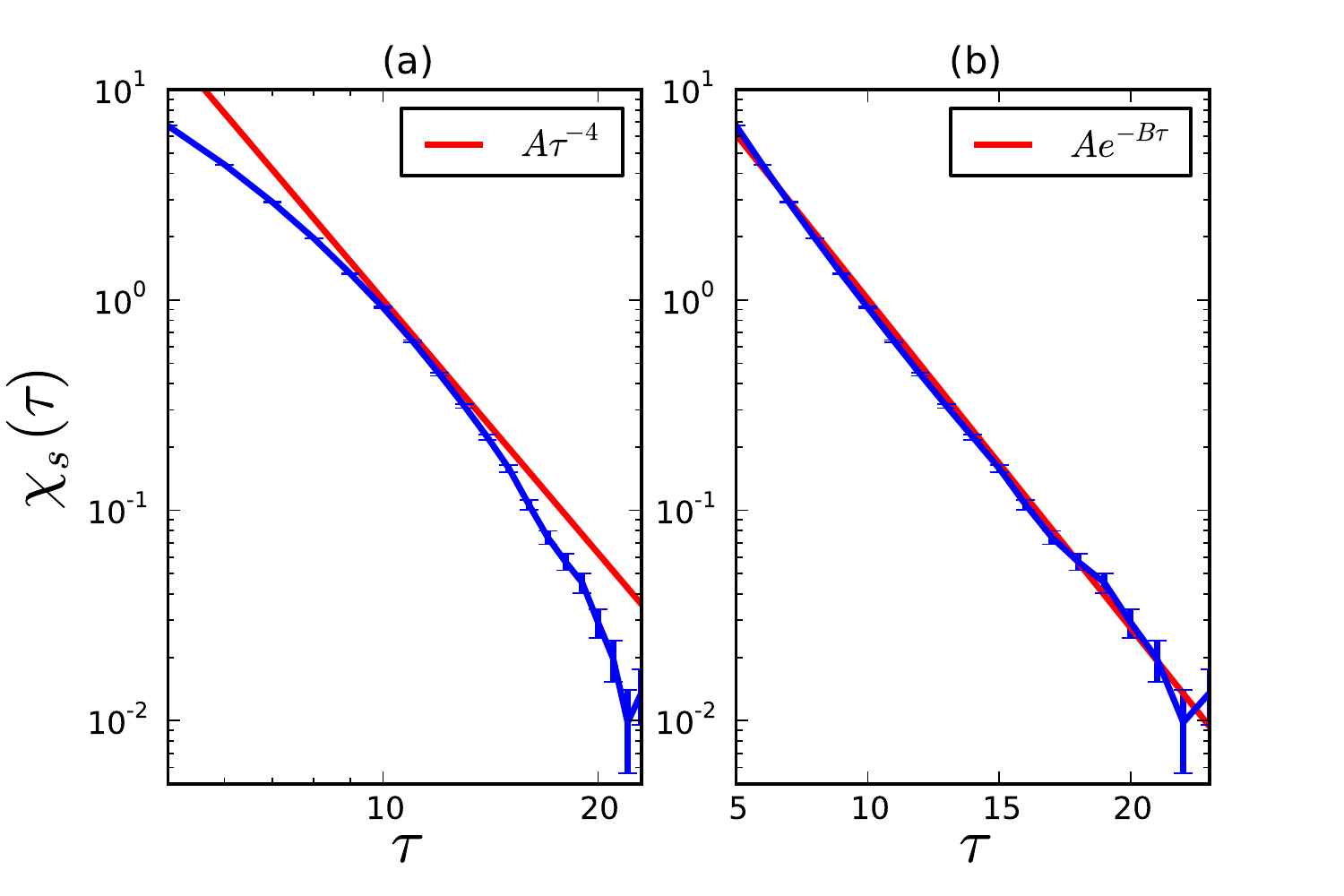}
\caption{$\chi_s(\tau)$ in the ordered phase for $N=2$, plotted on a log-log scale in panel (a) and a semi-log scale in panel (b).
The curve deviates significantly from the expected $1/\tau^4$ power law form. Instead, the curve fits better to an exponential decay
as in the disordered phase.}
\label{fig:tau4N2}
\end{figure}

In Fig.~\ref{fig:tau4} we present $\chi_s(\tau)$ on a log-log plot for $N=3,4$ in the disordered phase with the detuning parameter
$\delta g=0.1\times 10^{-2}$. For $N=3,4$ we indeed find agreement with the asymptotic behavior $\chi_s(\tau)\sim 1/\tau^{4}$ within
the error bars. In Fig.~\ref{fig:tau4N2} we present $\chi_s(\tau)$ for $N=2$ on a log-log plot and on a semi-lrog plot. Interestingly, for $N=2$
we do not find a conclusive asymptotic fall-off as  $1/\tau^4$.  Instead, the data fits better to an exponential decay, as in the
disordered phase. This indicates that the $\omega^3$ sub-gap spectral weight, if at all present, is small compared to the spectral weight contained in the
Higgs peak. Indeed we find excellent agreement between the large $\tau$ exponential decay rate and the value of $m\ns_H$ obtained
from the MaxEnt analysis, further supporting our results for the Higgs mass. We note that a $1/\tau^4$ power law behavior might be regained
at larger values of $\tau$, but this lies below the statistical inference of our data. 

Accurate determination of the scalar susceptibility at zero Matsubara frequency $\chi_s(i\omega=0)$ is crucial for this
analysis. Errors in $\chi_s(i\omega=0)$ translate into an overall vertical shift of $\chi_s(\tau)$. This error can dominate the value
of $\chi_s(\tau)$, especially at large $\tau$ where $\chi_s(\tau)$ is numerically small, and can lead to a bias in the power-law analysis.
Typically, $\chi_s(i\omega=0)$ is measured from a fluctuation relation
$\chi_s(i\omega=0)=\sum_{x,y,\tau}\avbra{\vphis_{x,y,\tau}\,\vphis_{\bf{0}}}-\avbra{\vphis_{\bf{0}}}^2$
and hence does not self-average \cite{Milchev_selfAv} upon increasing the system size. To overcome this difficulty we computed
$\chi_s(i\omega=0)$ using a direct numerical derivative  $\chi_s(i\omega=0)= -d\langle\vphis\,\rangle/d\mu$. To do so we evaluated
$\avbra{\phi^2}$ for a set of values of $\mu$ within a narrow range $[\mu-\Delta \mu,\mu+\Delta \mu]$  and extracted the derivative
by a polynomial fit in $\mu$. We found that this method reduced the error in $\chi_s(i\omega=0)$ by an order of magnitude and
significantly improved the power law decay analysis.

\section{Dynamical Conductivity}\label{secDC}

\begin{figure}[t!]
\includegraphics[width=0.475\textwidth]{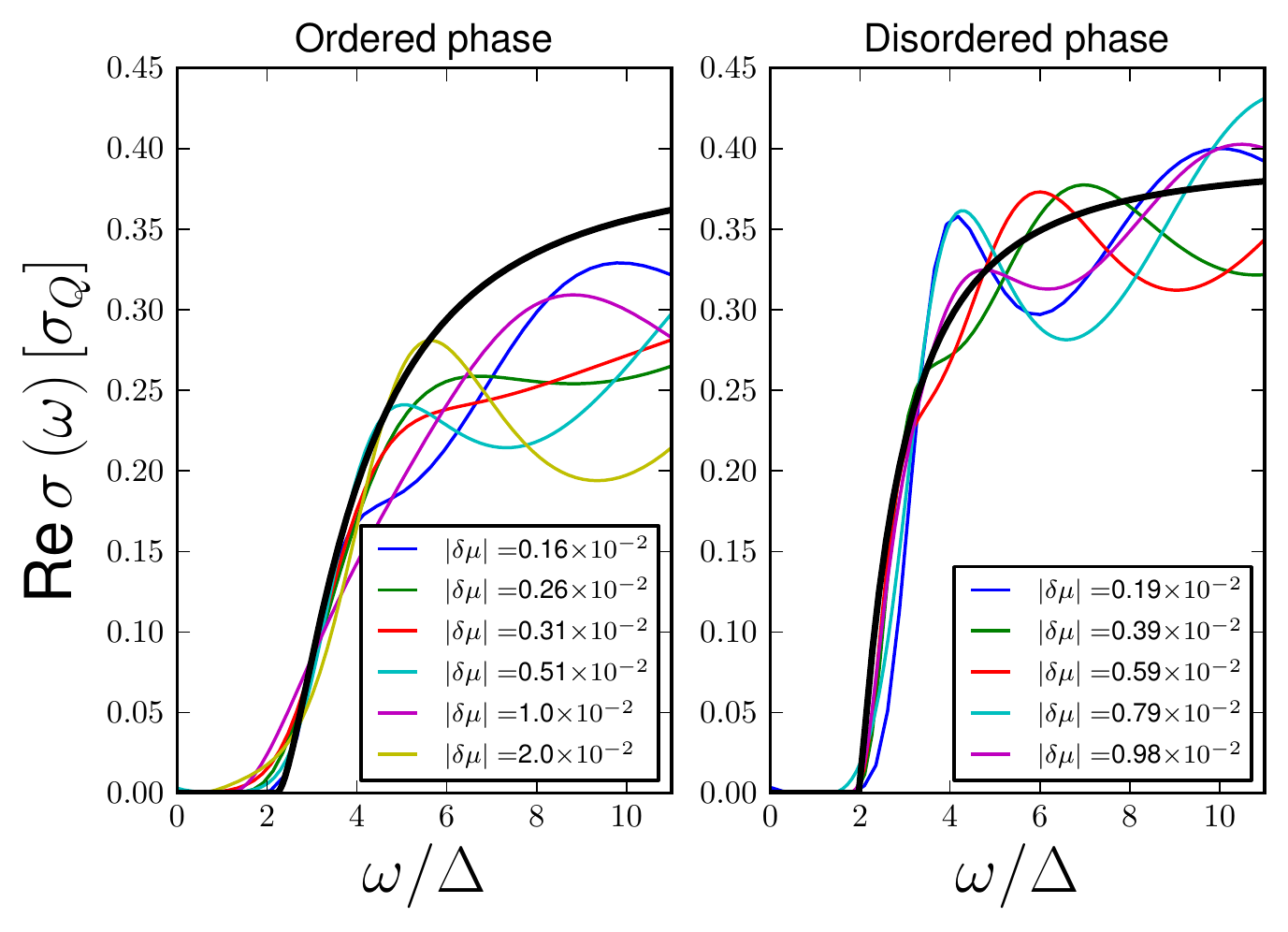}
\caption{The optical conductivity, $\Real\sigma(\omega)$ on the ordered and disordered phases. Curves are scaled by according to
Eq. \eqref{eq:condscal} for several values of the quantum tuning parameter $\delta g$ near the critical point. The solid black curves
show the analytic results from Refs. \onlinecite{Podolsky_visibility} and \onlinecite{damle_UniCond}}
\label{fig:CondBoth}
\end{figure}

In Fig. \ref{fig:CondBoth} we present the dynamical conductivity in the disordered and ordered phases. In both cases the frequency axis
$\omega$ is rescaled by $\Delta$, noting that there is no need for a vertical rescaling since the conductivity is a universal amplitude.
In both phases the curves collapse into single a universal shape, especially at low frequencies. The spectrum on the disordered side has
a clear gap-like behavior up to a threshold frequency $2\Delta$. Beyond this threshold, the spectrum rises sharply and saturates at a
universal value of $\sigma\ns_{\rm dis}(\omega \gg \Delta)\approx 0.35 (5)\,\sigQ\ns$, where $\sigQ\ns=e^{*2}/h$ is the quantum of
conductance.  These results should be compared with the line shape calculated diagrammatically in Ref. \onlinecite{damle_UniCond},
\beq
\sigma\ns_{+}(\omega) = 2\pi\sigQ \left({\omega^2-4 \Delta ^2\over 16\omega^2}\right) \Theta(\omega - 2\Delta)\ .
\label{eq:condQBE}
\eeq

Similarly, in the ordered phase, the dynamical conductivity grows rapidly starting at a threshold frequency $\approx 2 \Delta$,
and saturates at high frequency at a value $\sigma_{\rm ord}(\omega \gg \Delta)\approx 0.25 (5)\,\sigQ$.  A calculation to leading
order in weak coupling predicts\cite{Podolsky_visibility,Lindner_BadMetal} (see also appendix C)
\beq
\sigma\ns_{-}(\omega) = 2\pi \sigQ \left({\omega^2-m_H^2\over 4\omega^2}\right)^{\!\!2} \Theta(\omega - m\ns_H)\ .
\label{eq:condWC}
\eeq
In contrast to the disordered phase, there is a sub-gap component to the conductivity, owing to the gaplessness of the Goldstone mode(s).
This feature is first evident at two loop order in a perturbative calculation of the conductivity.  This was computed in Ref.
\onlinecite{Podolsky_visibility}, where it was found that the corresponding sub-threshold ($\omega < m\ns_H$) contribution to
$\sigma(\omega)$ is
\begin{align}
\sigma_-(\omega)\big|_{\omega < m\ns_H}&=\sigQ\cdot
{g m\ns_H\over2^{8}\pi}\Bigg\{{N-2\over N}\bigg({16\omega\over 15m\ns_H}+{32\omega^3\over 105 m_H^3}\bigg) + \nonumber \\
&\qquad + {3N-5\over N}{16 \omega^5\over 315 m_H^5}  + \ldots \Bigg\}+ {\cal O}(g^2)\ .
\end{align}
Remarkably, for $N=2$, the two leading order frequency terms in the sub-threshold conductivity vanish, resulting in a pronounced
pseudogap behavior.  Our numerical results appear to be qualitatively consistent with this analytic prediction.  However, the
coefficient of the leading $\omega^5$ term is small, given by $3.2\times 10^{-5}\,g/m_H^4$, and is not resolved within our
numerical accuracy.

For comparison, the analytic curves corresponding to Eqs.~(\ref{eq:condQBE}) and (\ref{eq:condWC}) are plotted in Fig. \ref{fig:CondBoth}.
The value of $m\ns_H$ was taken from the scalar susceptibility analysis\cite{gazit_fate} and $\Delta$ from the gap analysis. There is a
remarkable agreement between analytic and numerical curves especially at low frequencies.  It is important to notice that analytic curves are
presented without any fitting parameters (after setting $m\ns_H$ and $\Delta$). 

On general grounds, one expects the high frequency ($\omega\gg \Delta$) limit of the universal conductivity functions to be equal on both
ordered and disordered phases.  Here we find slightly different values, $\sigma\ns_{\rm dis}(\omega \gg \Delta)\approx 0.35 (5)\,\sigQ\ns$
and  $\sigma_{\rm ord}(\omega \gg \Delta)\approx 0.25 (5)\,\sigQ$, although there is significant spread which we attribute to limitations
of the analytic continuation.  This high frequency value should also match the universal conductivity in the quantum critical regime at
high frequencies ($\Delta=0$ and $\omega\gg T$).  Taking an average over both results, we estimate
$\sigma^*_c(\omega \gg T)\approx 0.3 (1)\,\sigQ\ns$.  This value should be compared with the value $\sigma^*_c=0.39\, \sigQ$ obtained in the large $N$ limit
in Ref.~\onlinecite{damle_UniCond}, and with $\sigma^*_c=0.251\,$\cite{Cha_unicond} obtained from leading correction in $1/N$. In addition, previous QMC simulations found $\sigma^*_c=0.33\, \sigQ$ \cite{Sorensen_UniCond} and $\sigma^*_c=0.285\, \sigQ$ \cite{Cha_unicond}.

\subsection{Charge-vortex duality of the dynamical conductivity}

The model in Eq.~\eqref{eq:ContModel} with $N=2$ describes relativistic bosons in 2+1 dimensions. This system has a dual representation
in terms of vortices \cite{fisher_duality}. Interestingly, the conductivity of the bosons is inversely proportional to the conductivity
of the vortices \cite{Cha_unicond}, such that
\begin{equation}
\sigB\ns(\omega)={\sigQ^2\over\sigV (\omega)}\ .
\label{eq:dualcond}
\end{equation}
Here $\sigB\ns = \sigma$ is the physical conductivity of the bosons, and $\sigV$ is the vortex conductivity in response to a dual electric
field, that is, a current of bosons. This relation is a direct consequence of the duality transformation and is therefore exact.

In the dual picture, the vortices interact with an inverse coupling constant.  This fact can be used to relate physical properties on
opposite sides of the transition.  This mapping is not exact due to the different interaction laws of bosons and vortices -- the bosons
have contact interactions whereas the vortices have long-ranged interactions.  This discrepancy prevents us from deriving exact results
from the duality relation, yet it can be used to construct approximate or qualitative results.   This relation was used in previous studies
to estimate the DC conductivity at the critical point where the vortices and bosons are self-dual, hence $\sigB\ns=\sigV\ns=\sigQ$. 
This simple argument, although not exact, gives the correct order of magnitude for the DC conductivity at the QCP.  

Here we ask whether this approach can be extended to the dynamical conductivity.  Duality maps the conductance of symmetric points on
both side of the transition:
\begin{equation}
\sigB\ns (\omega,-\delta g)=\sigV\ns(\omega,\delta g)\ .
\end{equation}
This relation, combined with Eq.~(\ref{eq:dualcond}), yields a relation between the optical conductivities on both sides of the transition:
\begin{equation}
\sigB\ns(\omega,-\delta g)={\sigQ^2\over \sigB(\omega,\delta g)}\ .
\label{eq:dualsigma}
\end{equation}
Here, $\sigB\ns(\omega,\delta g)$ is complex, containing both dissipative and reactive parts $\sigB\ns=\sigB'+i \sigB'',$ such that
\begin{align}
\sigB'(\omega,-\delta g)&={\sigQ^2\, \sigB'(\omega,\delta g)\over{\sigB'}^{\!2}(\omega,\delta g)+{\sigB''}^2(\omega,\delta g)}
\label{eq:dualreal}\\
\sigB''(\omega,-\delta g)&=-{\sigQ^2\, \sigB''(\omega,\delta g)\over{\sigB'}^{\!2}(\omega,\delta g)+{\sigB''}^2(\omega,\delta g)}\bvph\label{eq:dualimaginary}
\end{align} 
Note that duality flips the sign of the reactive component.  

Numerically we found it difficult to extract the reactive part of the conductivity.  The results for the analytic continuation were much
less numerically stable than for the dissipative component.  Yet, the numerics do provide some evidence for the duality.  According to
Eq.~(\ref{eq:dualreal}) one prediction of duality is that whenever the dissipative part vanishes for some frequency $\omega$ in one of the
phases, it must also vanish at the mirror point in the other phase.  This is indeed seen to be the case in  Fig.~\ref{fig:CondBoth},
where the threshold frequency of the dissipative part of the optical conductivity equals $\omega\ns_{\rm T} \sim 2\Delta$ on {\em both}
sides of the transition.  The presence of small subgap conductivity in the superfluid is a consequence of the inexactness of the duality.  

\begin{figure}[t!]
\includegraphics[width=0.5\textwidth]{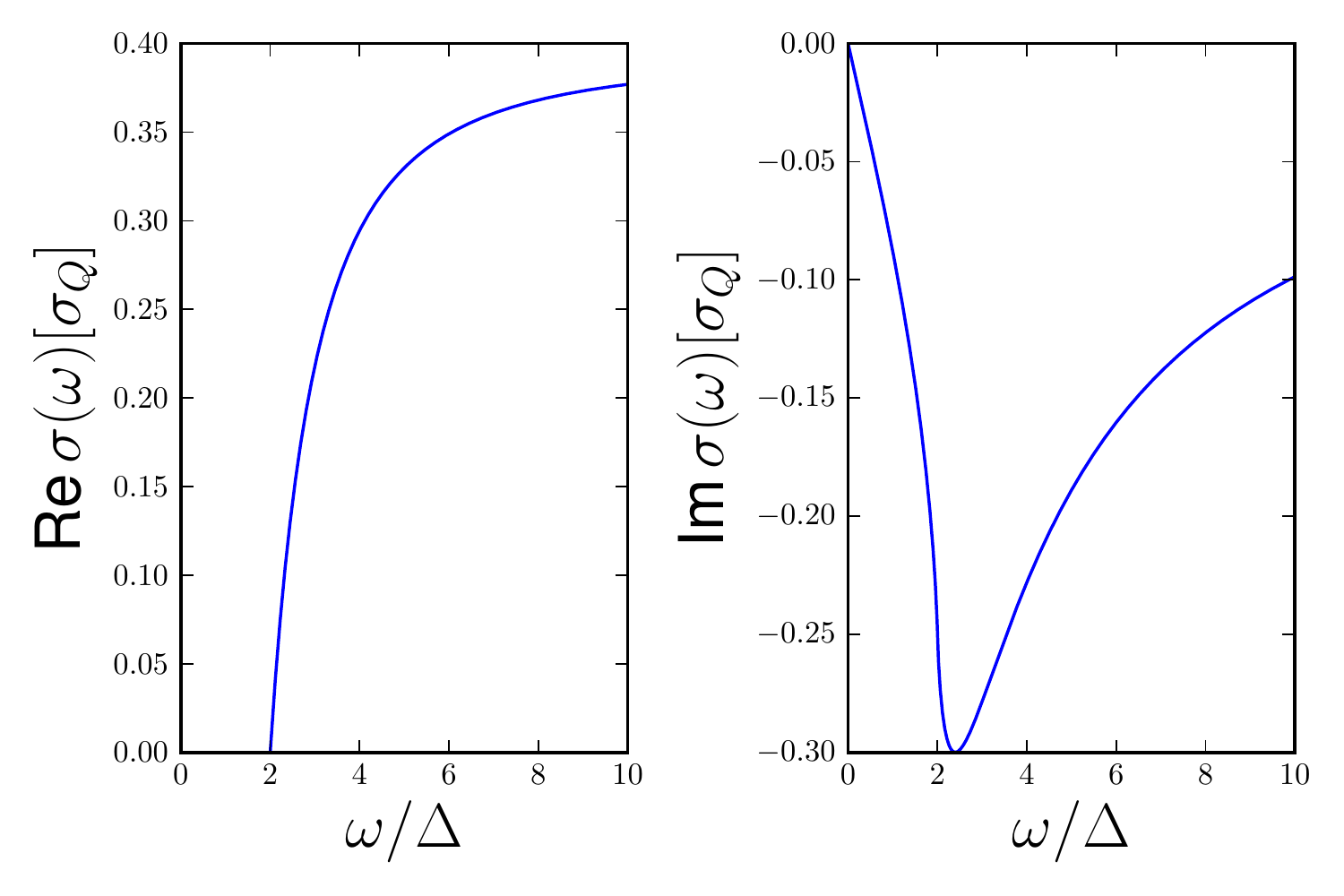}
\caption{The real and imaginary part of the optical conductivity in the disordered phase. Results are shown from a one loop calculation
in Appendix \ref{AppC}.}
\label{fig:CondDis}
\end{figure}

\begin{figure}[t!]
\includegraphics[width=0.5\textwidth]{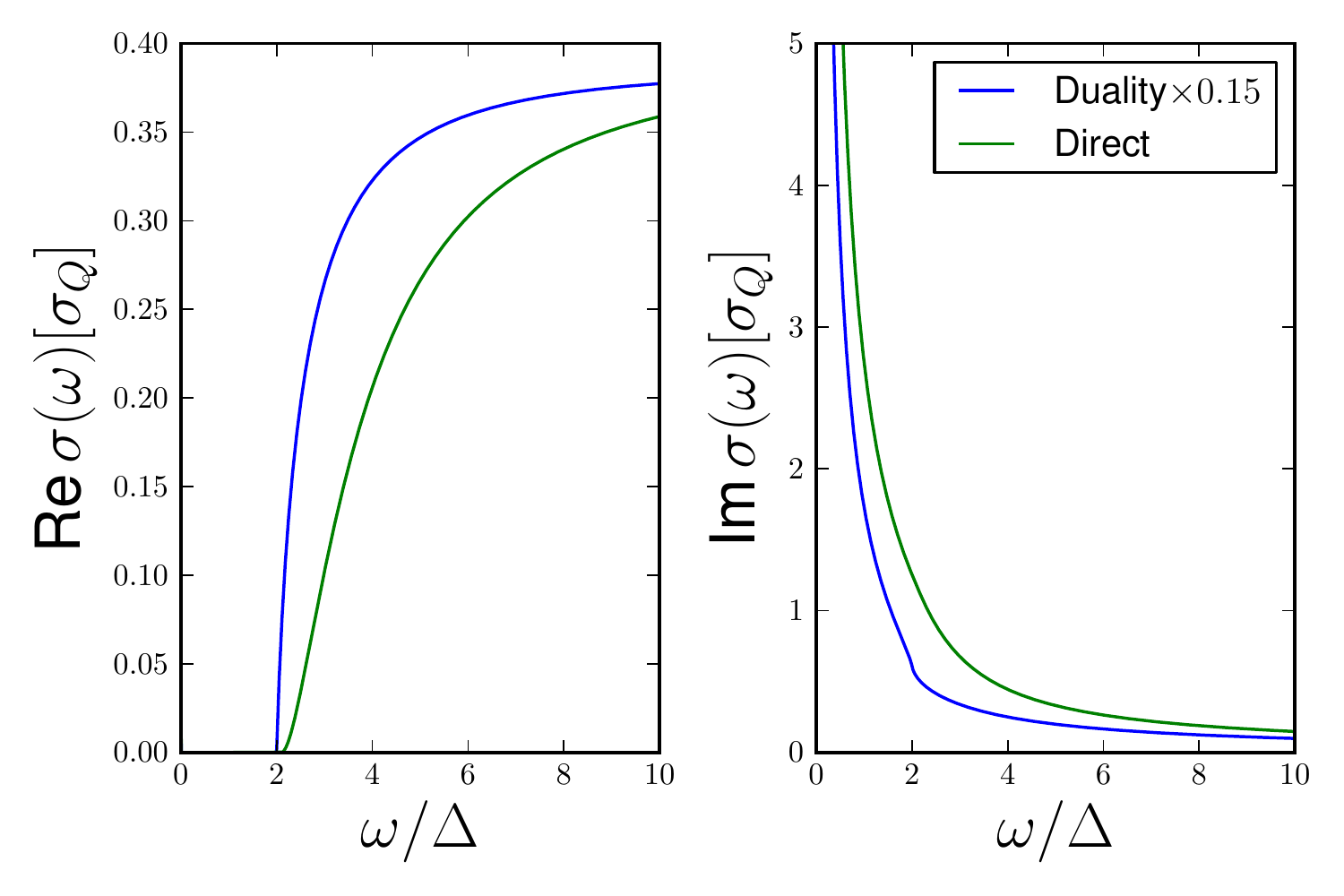}
\caption{The real and imaginary part of the optical conductivity in the ordered phase. The green curve displays the results of a one loop calculation carried in Appendix \ref{AppC}. The values of $m_H/\Delta$ and $\rho_s/\Delta$ were taken from the QMC simulation. The blue curve depicts the optical conductivity obtained from the duality relation in Eq. \eqref{eq:dualsigma}. This curve is multiplied by $0.15$ for comparison reasons.}
\label{fig:CondOrd}
\end{figure}

As an additional test of the duality, in Appendix C we present analytic calculations of the optical conductivity on both sides of the transition, to one loop order.  In Figs. \ref{fig:CondDis} and \ref{fig:CondOrd} we show the dynamical conductivity on the ordered and
disordered phase, respectively.  In order to use the same reference energy scale in both figures, we used the universal values
$m\ns_H/\Delta=2.1$ and $\rho_s/\Delta=0.44$ obtained numerically in  earlier parts of the analysis.  In Fig. \ref{fig:CondOrd} we also
depict the conductivity in the ordered phase, as obtained by applying the duality, Eq.~(\ref{eq:dualsigma}), to the conductivity in the
disordered phase.  As in the DC case, the overall scale of the conductivity has the right order of magnitude, set by $\sigQ$, but is
not quantitative.  However, the functional form of the conductivity is well captured by the duality.

Interestingly, duality makes a strong prediction on the reactive component of the conductivity at low frequencies.  A superfluid acts as a
perfect inductor at low frequencies, with admittance $\sigB''(\omega)=1/\omega L_{\rm ord}$, where the inductance is
$L_{\rm ord}=\hbar/(2\pi\sigQ \rho_s)$. \footnote{There is ambiguity in the literature regarding the sign convention of the reactive part
of the optical conductivity.  Here, positive/negative values reflect inductive/capacitive behavior.  This is opposite to the convention
often used in electric circuits.}  According to Eq.~(\ref{eq:dualimaginary}), using the fact that the dissipative part is negligible for
$\omega\ll \Delta$, this implies that at low frequencies the disordered phase behaves as a capacitor, with admittance is
$\sigB''(\omega)=-\omega C_{\rm dis}$, where the capacitance is $C_{\rm dis}= \sigQ\hbar/(2\pi\rho_s)$.  Physically, this capacitance measures
the polarizability of the bosons in the presence of an external electric field.  Furthermore, if the duality were exact, the universal
ratio between the capacitance in the disordered phase and the inductance in the ordered phase would be
$C_{\rm dis}/L_{\rm ord}=\sigQ^2$.
Indeed we find that at low frequencies the optical conductivity in the disordered phase, computed in Appendix C, rises linearly as
$\sigma_{\rm dis}''(\omega)=-2\pi\sigQ\times\hbar\omega/(24 \pi \Delta)+\mathcal{O}(\omega^2)$, that is, as a capacitor with capacitance
$C_{\rm dis}=2\pi \sigQ\times\hbar/(24 \pi \Delta)$.  This yields the ratio
\begin{equation}
{C_{\rm dis}\over L_{\rm ord}}={2\pi \rho_s\over 12 \Delta}\, \sigQ^2\approx 0.23\, \sigQ^2\ .
\end{equation}
where in the last equality we used $\rho_s/\Delta=0.44$ as obtained in Sec.~\ref{sec:stiffness}.

An illuminating way to understand the low frequency conductivity is through the dual vortex representation. In this representation the
effective field theory is given by a complex $\psi^4$ theory coupled to an electromagnetic gauge field\cite{stone_loops,Arovas_Dynaical}:
\begin{equation}
\begin{split}
{\cal S}&=\int d^3\!x \,\bigg\{ \big| (\pmu +ia_\mu)\psi \big|^2 + m^2 |\psi|^2 + \lambda |\psi|^4 \\
&\hskip 1.0in +{1\over 16\pi^2 K} \> F_{\mu\nu} F_{\mu\nu} \bigg\}\ .
\end{split}
\end{equation}
Here, the complex field $\psi$ is the vortex condensate order parameter field, $F_{\mu\nu}=\pmu a_\nu - \pnu a_\mu$, and $K$ is the coupling constant of
the bosons. The gauge field $a_\mu$ is related to the original boson 3-current by:
\begin{equation}
J_\mu={1\over 2\pi}\,\epsilon_{\mu\nu\lambda}\pnu a_\lambda
\end{equation}
Since the current is equal to the dual electric field $J_x(i\omega_m)=-\omega_m a_y/2\pi$, the conductivity is
\begin{equation}
\begin{split}
\sigma(i\omega_m)&= -{1\over\omega_m} \avbra{J_x(i\omega_m) \,J_x (-i\omega_m)}\\
&={\omega_m\over (2\pi)^2}\, \avbra{a_y(i\omega_m)\, a_y(-i\omega_m)}.
\end{split}
\end{equation}
In the disordered vortex phase, corresponding to the superfluid phase of the original bosons, the gauge field remains gapless with the
propagator in Feynman gauge:
\begin{equation}
\avbra{a_\mu\, a_\nu}={4\pi^2 K\over k^2}\,\delta\ns_{\mu\nu}\ ,
\end{equation}
hence the conductivity is $\sigma\ns_{\rm ord}(i\omega_m)=K/\omega_m$.  After analytic continuation and introducing physical units
$e^{*2}/\hbar = 2\pi \sigQ$, this becomes
\begin{equation}
\sigma_{\rm ord}(\omega)= 2\pi \sigQ\times {\rho_s\over \hbar}\bigg[{i\over \omega}+\pi \delta(\omega)\bigg]\ ,
\label{eq:sigmadualord} 
\end{equation}
where we have set $K=\rho_s$, its value  in the superfluid phase.

In the condensed vortex state, corresponding to the disordered phase of the original bosons, the field $\psi$ gets an expectation value
leading to a mass term for the gauge field through the Anderson-Higgs mechanism.  The effective action of the gauge field is then given by
a Proca action:
\begin{equation}
{\cal S}=\int \!\! d^3\!x \>\bigg\{ {1\over 16\pi^2 K}\,F_{\mu\nu} F_{\mu\nu}+{1\over 2}\rho\ns_v a_\mu^2\bigg\}\ .
\end{equation}
where we now take the vortex condensation density $\rho_{v}=2|\langle\psi\rangle|^2$. The gauge field propagator is
\beq
\avbra{a_\mu \, a_\nu}={4\pi^2 K \over k^2+M^2} \bigg(\delta_{\mu\nu}-{k_\mu k_\nu\over M^2}\bigg)\ ,
\eeq
where the gauge field mass $M$ is given by $M^2=4\pi^2 K \rho_v$. Note that, since the current is quadratic in boson operators,
this mass is related to the single-boson gap $\Delta$ by $M=2\Delta$.  Now the conductivity is given by
$\sigma\ns_{\rm dis}(i\omega_m)=\omega_m K/(\omega_m^2+M^2)$, which yields after analytic continuation
$\sigma\ns_{\rm dis}(\omega)= i \omega K/(\omega^2-M^2)$. At low frequencies $\omega\ll M$ this becomes, in physical units,
\begin{equation}
\sigma\ns_{\rm dis}(\omega)\approx - 2\pi i \sigQ \, {K\over M^2}\, \hbar\omega=-i\sigQ {\hbar \omega\over 2\pi\rho_v}\ .\label{eq:sigmadualdis}
\end{equation}
Combining the results from Eqs.~(\ref{eq:sigmadualord}) and (\ref{eq:sigmadualdis}) we obtain 
\begin{eqnarray}
{C_{\rm dis}\over L_{\rm ord}}={\rho_s \over \rho_v} \sigQ^2\ .
\end{eqnarray}
This gives a physical interpretation for the universal ratio $C/L$ as the ratio between the superfluid stiffness and the vortex
condensation density on opposite sides of the transition.

\subsection{Effect of Coulomb interactions}
Josephson junction arrays and granular superconducting films can often be described by charged lattice bosons\cite{mihlin2009}, which interact at long range via $e^2/r$  Coulomb interactions.
When Coulomb interactions are present, the O$(N)$ model Lagrangian should be augmented by a contribution
\begin{equation}
\Delta L=\int\!\! d^2\!x\ i n\,{\partial\varphi\over\partial\tau} + \half\!\int\!\! d^2\!x\!\!\int\!\!d^2\!x'
\> n(\Bx)\,{{e^*}^2\over |\Bx-\Bx'|}\,n(\Bx')
\end{equation}
where $\varphi$ is the phase of the order parameter.  We parameterize the $\vphi$ field in terms of longitudinal ($\sigma$)
and transverse ($\pi$) fluctuations:
\begin{equation}
\vphi=\big( \phi\ns_0+\sigma\,,\,\pi\big)\ ,
\end{equation}
where $\phi\ns_0\equiv \big|\langle\vphi\,\rangle\big|$.
To lowest order, we have $\varphi=\pi/\eta\sqrt{N}$, where  $\eta\equiv \phi\ns_0/\sqrt{N}$ is
proportional to the magnitude of the order parameter. Integrating out the density field $n(\Bx,\tau)$, we find that the $\pi$
propagator becomes
\begin{equation}
G\ns_{\pi\pi}(q)={1\over q_0^2+\Bq^2 + \alpha\, |\Bq| \,q_0^2}\ ,
\end{equation}
where $\alpha=\eta g \hbar v/\pi {e^*}^2$ and $v$ is the velocity (`speed of light') in the original O$(N)$ model.
This new $\pi$-field propagator has a 2D plasmon pole located at $q\ns_0=\sqrt{-|\Bq|/\alpha}$ for small $\Bq$.
Plugging this into the expression for the electromagnetic kernel, in Eq. (E1) of Ref. \onlinecite{Podolsky_visibility},
we find, to order $g^0$,
\begin{equation}
\sigma(\omega)=2\sigQ\,\Big({\alpha\over m_H}\Big)^{\!2}\> (\omega-m\ns_H)^4\>\Theta(\omega-m\ns_H)\ .
\end{equation}
Thus, the dynamical conductivity of two dimensional superconductors rises above the Higgs threshold with a modified power law  $\sigma(\omega)\propto (\omega-m\ns_H)^4$.

\section {Discussion and Summary}

In this work we studied the critical dynamical properties of O$(N)$-symmetric models with relativistic dynamics in two space dimensions.
In particular we computed the line shape of the scalar susceptibility and the optical conductivity on either side of the quantum phase
transition. Our results focus on properties that are universal in nature and are therefore relevant for many experimental realizations
of quantum phase transitions.

We showed that the scalar susceptibility, in the ordered phase, contains a clear resonance at the Higgs mass $m\ns_H$. By contrast, in
the disordered phase the scalar susceptibility has a threshold at $\omega=2\Delta$ with no conclusive evidence for a resonance above the
threshold. In addition we provide two universal dimensionless constants that characterize the dynamics: the ratio between the Higgs mass
and the single particle gap on mirror points across the transition, and the fidelity of the Higgs resonance. These predictions could be
tested by future, high resolution, experiments of the superfluid to Mott insulator transition in cold atomic lattices\cite{endres}.

It is important to note that, close to the critical point, the scalar susceptibility captures the low frequency behavior of a generic
experimental probe that couples to the order parameter amplitude and not to its direction \cite{gazit_fate}.

We have also presented results for the optical conductivity on both sides of the phase transition. In both cases we find a sharp rise of
the spectral function at $\omega\approx2\Delta$. The threshold frequency in the ordered phase can be associated with the Higgs mass
$m\ns_H$. This provides an independent estimate of the Higgs mass, one which agrees very well with the value obtained from the scalar
susceptibility analysis. In addition we have computed the high frequency ($\omega \gg T$) universal conductance
$\sigma_{\rm c}^*=0.3(\pm 0.1)\times\sigQ$. This value is with agreement with previous analytic calculations\cite{damle_UniCond}.
Unfortunately the low frequency (``hydrodynamic'') limit $\omega\ll T$ is not accessible in the QMC simulation, as was discussed in
Ref.~\onlinecite{damle_UniCond}. 

We observe an approximate  duality relation between the reactive components of the conductivity in both phases. The
ordered (disordered) phase displays an inductive (capacitive) behavior, where the ratio $C/L$ between the capacitance $C$ and
inductance $L$ is found to be universal. To one loop order $C/L=0.23\,\sigQ^2$. Furthermore, we show that the dual
vortex representation predicts an interesting physical interpretation to the admittance ratio,  $C/L=\rho\ns_s\sigQ^2/\rho\ns_v$, where
$\rho_s$ is the superfluid stiffness and $\rho_v$ is the vortex condensation density in the two phases. Both impedances are
can be computed directly from a Monte Carlo simulation without analytic continuation.  We intend to do
this in a future study. Finally we have shown that for charged system with Coulomb interaction the power law of the spectral rise above
the threshold changes from 2 to 4. 

We hope that our results will motivate measurements of the optical conductivity in cold atoms by optical lattice phase modulation,
as was suggested in Ref. \onlinecite{Giamarchi_Phase}. Such experiments could accurately measure the universal optical conductivity
near the QCP and even the universal resistivity right at the critical point. Our analysis may also shed light on recent experiments
on the superconductor to insulator transition in granular superconductors\cite{ShermanTHz}. In this context will be interesting to
extend these calculations to systems with varying degrees of disorder.

{\it Note added\/}: After this work was completed, we learned of similar quantum Monte Carlo results by Witczak-Krempa, Sorensen, and Sachdev\cite{Witczak_ads},
who find a critical conductivity $\sigma(\omega/T\to\infty)=0.32\,\sigQ$ at the critical point, in good agreement with our estimate of
$0.3\,\sigQ$.  We thank William Witczak-Krempa for informing us of their work, and for some additional relevant references.

\acknowledgements

We thank Efrat Shimshoni, Aviad Frydman, Daniel Sherman and Nandini Trivedi for useful discussions. We thank the
Aspen Center for Physics, supported by NSF-PHY-1066293, for its hospitality. We acknowledge support from the Israel  Science Foundation,
the U.S.-Israel Binational Science Foundation, and the European Union
under grant agreement no. 276923 -- MC-MOTIPROX. SG received
support from a Clore Foundation Fellowship and wishes to thank Maya Epler.
DPA gratefully acknowledges support from NSF grant DMR-1007028.

\appendix
\section{Worm algorithm for O$(N)$ models}\label{AppA}

We present a novel QMC algorithm for O$(N)$ lattice models Eq.~\eqref{eq:latModel}. The algorithm is based on the worm algorithm 
\cite{prokofev_worm_2001} extending it for general O$(N>2)$ models. The first step is to expand Eq.~\eqref{eq:latModel} in strong coupling:
\begin{equation}
\CZ=
\int \!\mathcal{D}\vphi \ 
\prod_b\prod_{\alpha}\sum_{n_b^\alpha} {1\over n^\alpha_b!}
\big(\phi_{i\ns_b}^\alpha\phi_{i'_b}^\alpha\big)^{n_b^{\alpha}} \prod_j e^{-V(|\vphi\ns_j|^2)}
\end{equation}
with ${\cal D}\vphi\equiv \prod_i d^N\!\phi\ns_i$. Here $\{b\}$ represent the set of all lattice bonds, the site $i\ns_b$ is linked to
the site $i'_b$ through the bond $b$, the index $\alpha\in\{1,\ldots,N\}$ labels the $N$ components of each $\vphi\ns_i$, and
$V(s)= \mu s + g s^2 $ is the local on-site interaction. Next we integrate out the fields $\vec{\phi_i}$. This can be achieved by noting
that now the functional integral factorizes into a product of {\it single site integrals\/}, such that
\begin{equation}
\CZ=\sum_{\cbra{n_b^\alpha}}\prod_{b,\alpha} { 1 \over n_b^\alpha !}\prod_i W\big(\{k_i^\alpha\}\big)\ .
\label{eq:loopPartFunc}
\end{equation}
Where we define $k_i^\alpha=\sum_{b(i)}' n^\alpha_b$ as the sum over all bonds $b$ emanating from site $i$.
The single site weight is then
\begin{equation}
W\big(\{k_i^\alpha\}\big)  =\int \!\! d^N\! \phi_i \prod_\alpha \bra{\phi_i^\alpha}^{k_i^\alpha}  e^{-V(|\vphi|^2)}\ .
\label{weight}
\end{equation}
We may write
\begin{align}
W\big(\{k_i^\alpha\}\big) &= \int\!\!d^N\!\phi\ns_i\!\int\limits_0^\infty \!\! ds \, e^{-V(s)}\, \delta\big(s-|\vphi_i|^2\big)
\prod_\alpha (\phi^\alpha_i)^{k^\alpha_i} \nonumber \\
&= {1\over 2\pi}\!\int\limits_0^\infty\! ds\> e^{-V(s)}\!\! \int\limits_{-\infty}^\infty \!\!\!d\lambda\>e^{i\lambda s}
\prod_\alpha {\cal I}(k^\alpha_i)\ ,\label{Weqn}
\end{align}
where
\begin{equation}
\begin{split}
{\cal I}(k^\alpha_i)&=
\int\limits_{-\infty}^\infty\!\!\!d\phi^\alpha_i\,e^{-i\lambda (\phi^\alpha_i)^2}\,(\phi^\alpha_i)^{k^\alpha_i}\\
&= (i\lambda)^{-(k^\alpha_i+1)/2}\,\Gamma\big(\half+\half k^\alpha_i\big)\,\delta\ns_{k^\alpha_i,{\rm even}}\ .
\end{split}
\end{equation}
We now encounter the integral
\begin{equation}
\int\limits_{-\infty}^\infty\!\!\! d\lambda\> e^{i\lambda s}\,(i\lambda)^{-J}=2\,s^{J-1}\,\Gamma\big(1-J\big)\,\sin(\pi J)\ ,
\end{equation}
where $J=\half(N+K\ns_i)$, and $K\ns_i=\sum_\alpha k^\alpha_i$.  The above integral converges only if $0 <{\rm Re} J < 1$, however
our initial expression in Eq. \ref{weight} is clearly convergent for all possible values of $J$, which licenses us to analytically
continue the above expression, using the identity $\Gamma(J)\,\Gamma(1-J)=\pi/\sin(\pi J)$.  We then obtain
\begin{equation}
W\big(\{k_i^\alpha\}\big) = Q\big(\half N+\half K\ns_i\big)\prod_\alpha \Gamma\big(\half+\half k^\alpha_i\big) \,
\delta\ns_{k^\alpha_i,{\rm even}}\ ,
\label{eq:wormweight}
\end{equation}
with
\begin{equation}
Q(J)={1\over\Gamma(J)}\int\limits_0^\infty\!\!ds\> e^{-V(s)}\,s^{J-1}\ .
\label{Qeqn}
\end{equation}
The one-dimensional integrals $Q(J)$ can be evaluated numerically to high precision and tabulated prior to the QMC simulation.
In this representation the partition function sum runs over all integer values of the bond's strength $n_b^\alpha$, replacing the 
$\vphi_i$ field integrations. The sum is restricted only to closed path loops due to constraint $\delta_{k_i^\alpha,{\rm even}}$.

The updating procedure closely follows the worm algorithm, considering an extended partition function:
\beq
\CZ_G=\sum_{i,j} \avbra{\phi_i^\alpha\phi_j^\alpha}
\eeq
The fields insertion $\phi_i^\alpha\phi_j^\alpha$  breaks the closed path condition by adding a single open loop. The open loop's head 
is located at $i$ and its tail at $j$.

For simplicity we choose the open loop to be one of the flavors $\alpha$. The updating procedure consists out of two elementary steps. The first move is a shift move in which we move the worm's head to one of the neighboring sites connected with the bond $b$. During the move we either increase or decrease the bond's strength $n_b^\alpha$. The second move is a jump move, which is relevant only for closed loops where the head and the tail are located in the same site. We choose one of the lattice sites and jump with the head tail pair to that site. The QMC acceptance ratios can be easily derived from Eq.~\eqref{eq:wormweight} and Eq.~\eqref{eq:loopPartFunc} similarly to the argument in
Ref. \onlinecite{prokofev_worm_2001}.

We tested the correctness of our numerical implementation by comparing with previous QMC simulation and to analytic results of the Gaussian model limit of Eq.~\eqref{eq:latModel} ($g=0$). The results agree within the statistical errors.

We also provide an explicit expression for the sampling of the scalar susceptibility in the closed path representation. The operator
insertion $\vphis_i$ effectively introduces a factor of $s$ to the integrand in Eq. \ref{Weqn}, in which case Eq. \ref{Qeqn} is replaced
by $J\ns_i\,Q(J\ns_i+1)$.  Inserting $(\vphis_i)^2$ introduces a factor of $s^2$ and results in $J\ns_i (J\ns_i+1)\,Q(J\ns_i+2)$.
Thus, the insertion $\vphis_i\vphis_j$ yields
\begin{equation}
\begin{split}
\avbra{\vphis_i\vphis_j}&=\bigg\langle {J\ns_i J\ns_j\, Q(J\ns_i+1)\,Q(J\ns_j+1)\over Q(J\ns_i)\,Q(J\ns_j)}\bigg\rangle \quad (i\ne j)\\
&=\bigg\langle{ J\ns_i (J\ns_i+1) \,Q(J\ns_i+2)\over Q(J\ns_i)}\bigg\rangle \qquad\quad (i=j)
\end{split}
\end{equation}

\section{Analytic continuation of imaginary time QMC data}\label{AppB}

\subsection{General Formulation}

We use imaginary time action Eq.~\eqref{eq:latModel} in the QMC simulations in order for the QMC weights to be real and positive,
avoiding the dynamical sign problem. The real frequency, dissipative response function $A(\nu)$, can be obtained by numerical analytic continuation~\cite{jarrell_bayesian_1996}, which amounts to inverting
the equation,
\beq
\CG (i\omega_m)=\int\limits_0^\infty \!\!{d \nu\over\pi} {2 \nu\over \omega_m^2 +\nu^2}\>A(\nu)\ .
\label{eq:contAC}
\eeq

The kernel
\beq
K(m,\nu)={1 \over \pi}\cdot{2 \nu \over \omega_m^2 +\nu^2},
\label{Kernel}
\eeq
needs to be inverted in order to formally obtain,
\beq
A(\nu)=K^{-1}\CG (i\omega_m)\ .
\eeq
Unfortunately $K$  is an ill conditioned operator. The inversion is extremely sensitive to inevitable statistical noise in $\CG$.

\begin{figure}[!t]
\begin{center}
\includegraphics[width=0.475\textwidth,angle=0]{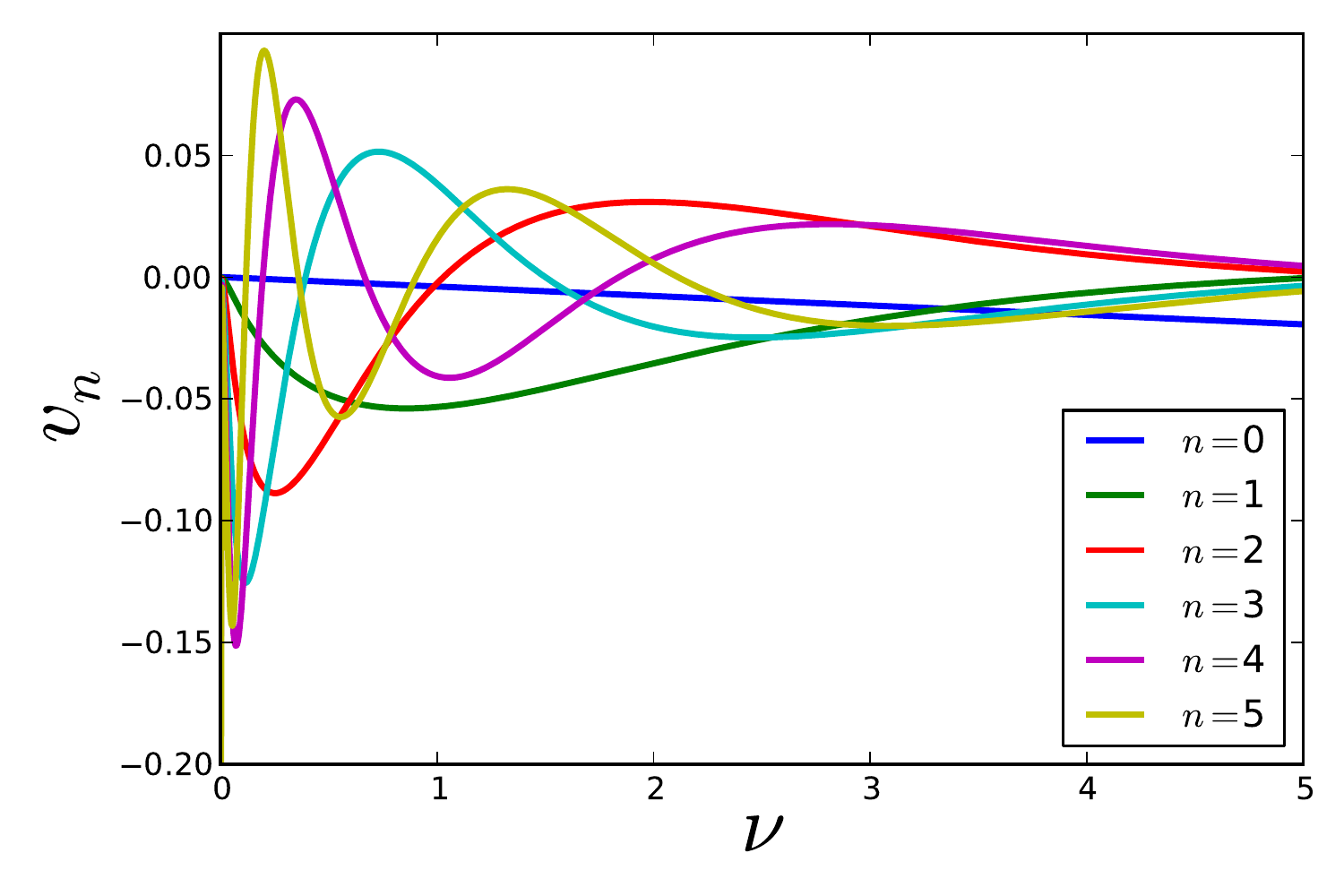}
\caption{The first five vectors $v_n(\nu)$ corresponding to the largest singular values in $W$.}
\label{V4}
\end{center}
\end{figure}

The stability of the inversion problem can be analyzed by the Singular Value Decomposition (SVD) 
\beq
K=UWV^T
\label{svd}
\eeq
where $U$ and $V$ are unitary matrices whose rows are the eigenvectors  $\langle u_n|$, and $\langle v_n|$.   The first five eigenvectors $v_n(\nu)$ are plotted in Fig. \ref{V4}.
$W$ is diagonal with real, non-negative SVD eigenvalues $w_n$. These are plotted on a logarithmic scale as a function of $n$  in Fig. (\ref{fig:toymodel}).
$W$ has up to ${\cal N}$ non zero singular values, where ${\cal N}$ is the number of QMC data points.

From Eq.~(\ref{svd}),  the  pseudo inversion of $K$ is given by
\beq
	\bar{K}^{-1} =  V \bar{W}^{-1}  U^{T}  \ .
	\label{eq:pseudoinv}
\eeq
Here $\bar{W}$  is a square diagonal matrix which contains only the non zero eigenvalues $w_n\ne 0$.

The SVD eigenvalues $w_n$ can be calculated by diagonalizing the Hermitian matrix $(K K^\dagger)_{ij}$:
\begin{equation}
\begin{split}
(K K^\dagger)_{ij}&=\int\limits_{-\infty}^\infty\!{d\omega\over 2\pi}\>{\omega^2 \over (\omega_i^2 + \omega^2) (\omega_j^2 +\omega^2) } \\
&={1\over 2 \big(|\omega_i|+ |\omega_j|\big)} = {\beta \over 4\pi } {1\over |i|+|j|} \ .
\label{eq:AC1}
\end{split}
\end{equation}
Since $\CG (\tau)$ is real, $\CG _n= \CG _{-n}$, and by projecting out the zero mode, we may restrict both $i$ and $j$ to be  positive integers in Eq. (\ref{eq:AC1}).

Matrices of the form
\begin{equation}
H\ns_{ij}(\tau,\theta)={\tau^{i+j}\over i+j+\theta}\quad,\quad (i,j)\in\{0,\ldots, N\}   .
\end{equation}
are known as Hilbert matrices.  We are interested in the case of $\tau=1$ and $\theta=2$.
An exact bound on the  dependence of the smallest eigenvalue on the matrix size was obtained by
Ref. \onlinecite{hilbert_asym},
\begin{equation}
\begin{split}
w_{\rm min}^{(N)} &\sim \kappa\, \sqrt{N}\, \Big(1+\sqrt{2}\Big)^{-4N} \!\times \big(1+ o(1)\big)  , \\
\ln\!\big(w_{\rm min}^{(N)} \big)  &\sim   - 3.52549\, N  + 0.5 \ln N + 0.7909  ,
\end{split}
\end{equation}
with
\begin{equation}
\kappa={2^{15/4}\,\pi^{3/2}\over\big(1+\sqrt{2}\big)^4}=2.205385\ldots
\end{equation}
As we see, the minimal eigenvalue decreases faster than exponentially with $N$, which is consistent with the behavior found numerically in Fig.~(\ref{fig:toymodel}).

\subsection{Pseudo-inversion by truncated SVD}

In practice, the noisy QMC data, called $\tilde{\CG}$ can be decomposed as
\be
\tilde{\CG}= \CG^{\rm sig} + \xi
\label{cg}
\ee
where $\CG^{\rm sig}$ is the true signal, and $\xi$ is a random noise.
The noise interferes with the numerical inversion of $\CG^{\rm sig}$.
To see this,  the  data $\tilde{\CG}$ is projected onto the eigenvectors $u_n$, which yields the real numbers 
\begin{equation}
\begin{split}
\tilde{p}_n =  \langle \,u_n\,|\, \tilde{\CG}\, \rangle  &=  \langle \,u_n\,|\, \CG^{\rm sig}\,\rangle +  \langle \,u_n\,|\, \xi\,\rangle \\
&\equiv p^{\rm sig}_n + \xi_n  .
\end{split}
\end{equation}

The pseudo inversion Eq.~\eqref{eq:pseudoinv} yields
\be
A(\nu) = \sum_n  {\tilde{p}_n  \over w_n}\> v_n(\nu) =   \sum_n \left( { p^{\rm sig}_n  \over w_n}\> v_n(\nu)+  {\xi_n  \over w_n}\> v_n(\nu)\right)  .
\ee
Since $\CG^{\rm sig}$ is the analytic continuation of a normalizable function,  $\sum_n |p^{\rm sig}_n/w_n|^2 $ must converge. This implies that  $|p^{\rm sig}_n|< w_n$ at large $n$.  On the other hand $\xi_n$  is not the analytic continuation of a normalizable function, and therefore is not necessarily bounded by 
$w_n$. For white noise, $\xi_n$ are random numbers whose variance is independent on $n$.

Therefore, one can readily identify a {\em breakpoint},  $n^*$,  which for $n<n^*$, $\tilde{p}_n\approx p^{\rm sig}_n$,  and for $n>n^*$,   $\tilde{p}_n\approx \xi_n$. The breakpoint serves to truncate the inversion and eliminate the dominance of noise terms. It can also allow an estimate of the truncation error.

Let us illustrate this procedure  by a  test model,
\beq
A^{\rm model}(\nu)=\nu^3\left(e^{-(\nu-\Delta)^2}+e^{-(\nu+\Delta)^2}\right) .
\label{eq:toymodel}
\eeq.

\begin{figure}[t!]
\includegraphics[width=0.475\textwidth]{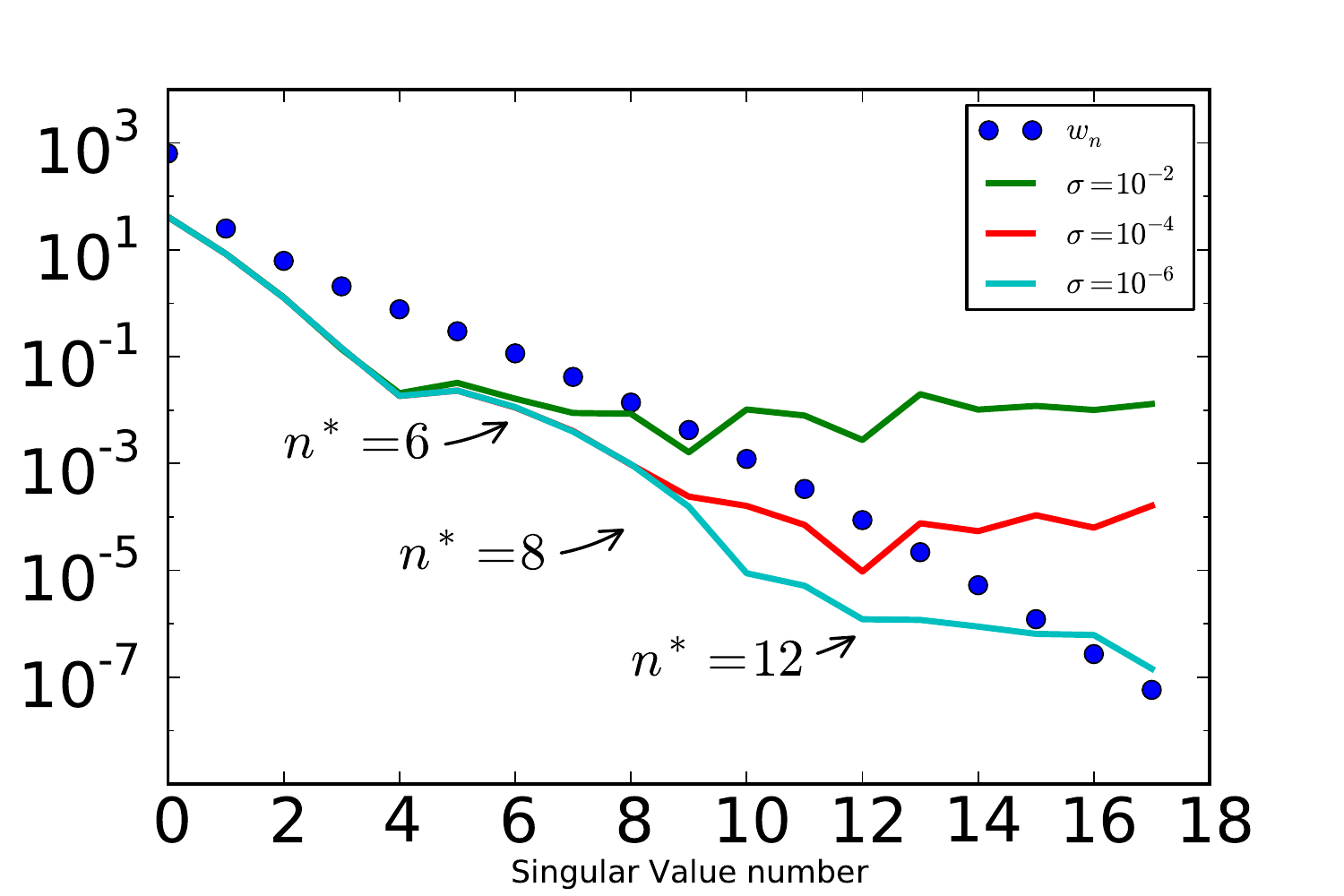}
\caption{SVD analysis of the numerical analytical continuation.  $n$ labels the SVD eigenmodes.
The filled circles are the rapidly decreasing SVD eigenvalues of $K$, denoted $w_n$. Magnitudes of projections of noisy data, for the  test model, Eq.~(\ref{eq:toymodel}), are denoted by  $|\tilde{p}_n| $. $\sigma$ is the variance of the artificial noise added to the Matsubara data.
The breakpoints $n^*$ denotes the mode index where noise dominates the signal, and the projections start to flatten.  The values of $n^*$increase when the noise level decreases.}
\label{fig:toymodel}
\end{figure}

If Fig.~\ref{fig:toymodel}  Eq.~\eqref{eq:toymodel} to the $u_n$ basis. $w_n$
and  $|p^{\rm sig}_n|$ rapidly decay, as expected from the Riemann-Lebesgue lemma for a smooth spectral function. We add an artificial white noise with increasing variance  $\sigma$. As expected, the (approximately) exponential  decay of ${\tilde p_n}$ stops
abruptly at $n^*$, where  $|\tilde{p}_{n^*}| \approx |\xi_{n^*}|$.  

As seen in Fig.~\ref{fig:toymodel}, the breakpoints $n^*$  are chosen where the curves average slope flattens abruptly.  $n^*$  increase as the noise is reduced.

A truncated SVD inversion provides  a controlled approximation for the spectral function:
\beq
\tilde{A}^{\rm svd}(\nu)\simeq  \sum_{n=1}^{n_{\rm svd}}  { \tilde{p}_n  \over w_n}\> v_n(\nu),
\label{trunc}
\eeq
The modes higher than $n_{\rm svd}$  are discarded because their coefficients, (which  only contribute random noise to  the spectral function), blow  up exponentially with $n$.
If we know the  bound on the signal's  convergence rate  $|p^{\rm sig} / w_n|^2 <  c ~e^{-\alpha n}$, we can estimate the  error in the norm as
\be
||\delta\tilde{A}||^2  =  \sum_{n=n_{\rm svd}+1}^{\cal N}  \left|{ p^{\rm sig}_n  \over w_n}\right|^2 <  { \tilde{p}_{n_{\rm svd}}^2   \over  \alpha w^2_{n_{\rm svd}} } .
\label{error}
\ee
Thus, the smaller the noise level, the larger $n_{\rm svd}$ and therefore the smaller the error in the spectral function, Eq.~(\ref{error}). 

\begin{figure}[t!]
\includegraphics[width=0.475\textwidth]{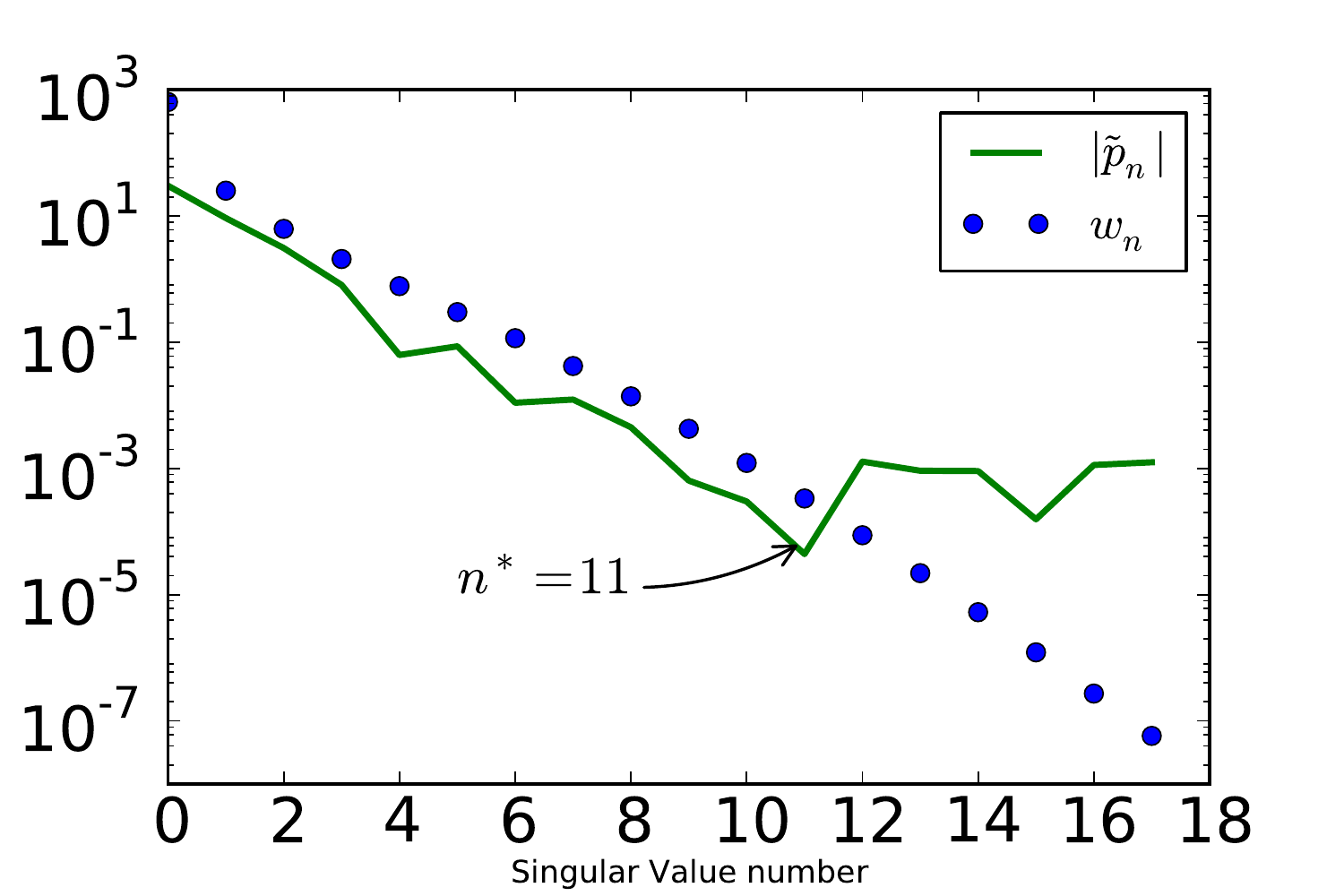}
\caption{Comparison between the projections $p_n$ (sold likes) and the singular values of the kernel $w_n$ (circles) for the high quality QMC data for the O(2) model. The linear system size is $L=120$, and coupling constant
is  $\delta g=1.17\%$. We see that
due to the effect of the noise,  $\tilde{p}_n$ flattens at the  breakpoint at $n^* \approx 11$.}
\label{fig:singvals}
\end{figure}

In Fig. \ref{fig:singvals} we show the SVD analysis of the QMC data for the real O(2) model.  The projections
$\tilde{p}_n$ flatten roughly at $n^* \approx 11$, as they behave in the test model in Fig~(\ref{fig:toymodel}).  

$\tilde{A}^{\rm svd}$ can exhibit spurious oscillations due to the missing modes $ \{v_n(\nu),~ n>n_{\rm svd} \}$. This effect, which is part of the error $||\delta\tilde{A}||^2$, is similar to spurious oscillations obtained by a truncated inverse Fourier transform.  
In cases where it is known that $A(\nu)>0$, (as for the scalar susceptibility and real conductivity), the SVD truncation can produce unsightly negative regions.

\begin{figure}[t!]
\includegraphics[width=0.475\textwidth]{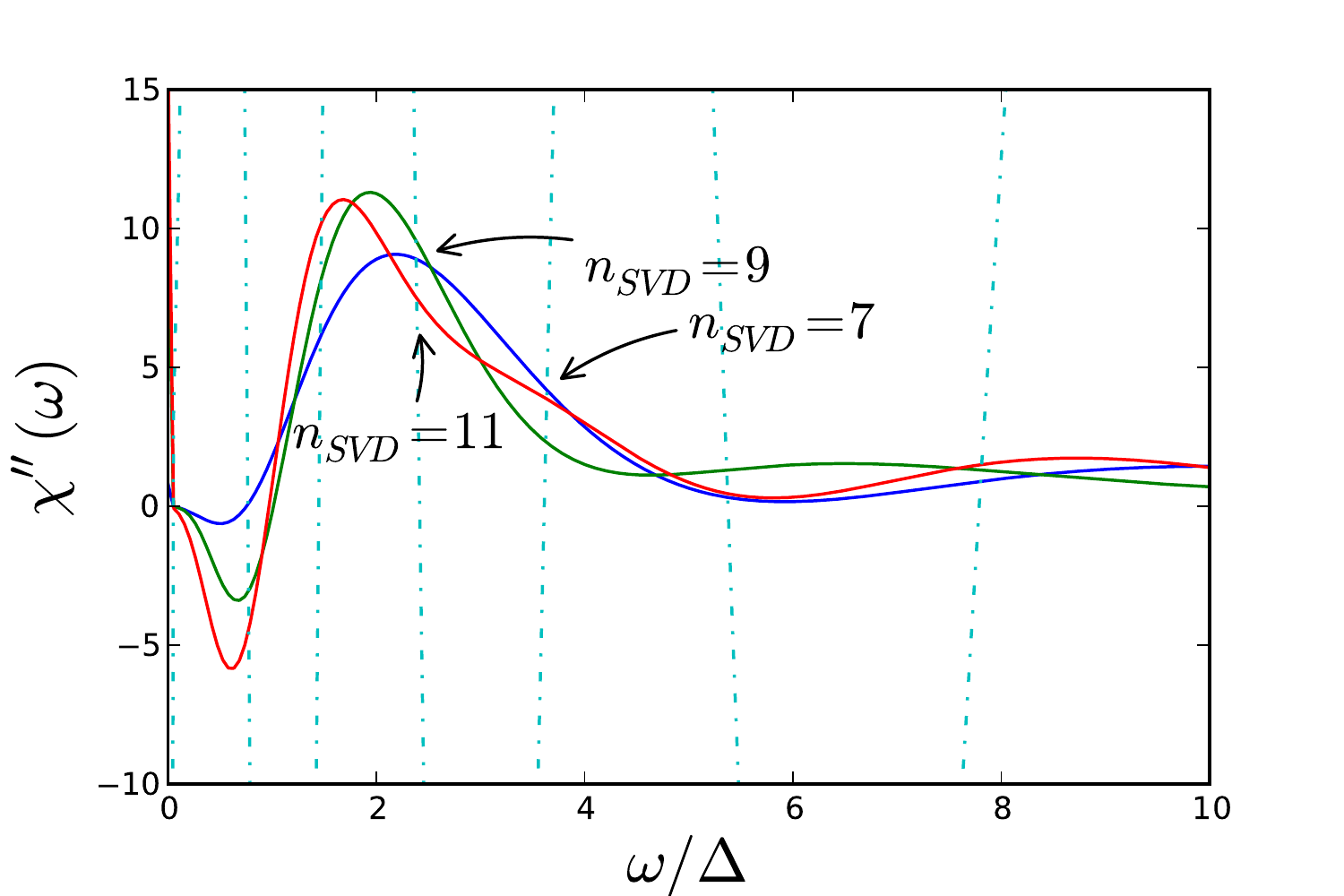}
\caption{Analytic continuation obtained from the first $n$ singular values, for the QMC data of Fig.~\ref{fig:singvals}. 
We see that spectral functions converge  to until $n\le 11$, in agreement with assigning $n^*=11$ where  $\tilde{p}_n$ starts to flatten in Fig.~\ref{fig:singvals}. 
For $n_{\rm svd}=12$, (dashed line) the condition $\tilde{p}_n<w_n$ is violated, and the resulting spectral function wildly differs from the converged function, since it is dominated by random amplified errors.
}
\label{fig:svdchop}
\end{figure}

 In Fig.~\ref{fig:svdchop} 
 we plot the $\tilde{A}^{\rm svd}(\nu)$  for increasing values of  $n_{\rm svd}$. We see that indeed the reconstructed solution converges as we increase $n$ and remains stable
up to $n\approx11$, which is where we locate the breakpoint in the SVD analysis. For $n=12$, the inverted errors dominate the spectral function, which  yields a wildly erroneous result.

\subsection{Maximum Entropy and other regularizations}
\label{sec:MaxEnt}

The QMC simulation produces noisy variables $\CG (i\omega_m)$, whose  covariance matrix is defined as
\beq
\Sigma^{-1}=\avbra{\CG (i\omega_m)\,\CG (i\omega_n)}.
\eeq
A condition for the inverted spectral  function is that
\beq
\chi^2=(\CG -K A)^T\, \Sigma\, (\CG -K A) \approx {\cal N}  ,
\eeq
where ${\cal N}$ is the number of data points.


As we have seen before, since $K$ has very small SVD eigenvalues, there is a large family of functions $A(\nu)$ which have the same  $\chi^2 / \mathcal{N} \approx 1$. The SVD truncation is one way to choose among these functions, but the result may have spurious oscillations, and turn negative in some regions.  To improve on this approximation one needs to impose extra conditions on $A(\nu)$, which amounts to extrapolation of
Eq.~(\ref{trunc}) to include higher SVD modes.
A common approach,  which ensures positivity, is  to introduce  a cost functional $f(A)$, and to variationally minimize  
\beq
Q = \frac{1}{2} \chi^2+\lambda f(A)   
\eeq
with respect to $A$.  This minimization lifts the degeneracy in $\chi^2$, and depends critically on the choice of  $\lambda$.  
$\lambda$ can be chosen by the L-curve method \cite{Lcurve}, which  is analogous to the determination of the breakpoint   $n^*$  described above.

Two cost functions are commonly used.  (1) The  `Maximum Entropy'  (MaxEnt)~\cite{jarrell_bayesian_1996},  
\be
f^{\rm MaxEnt}(A)=-\sum_i A(\nu_i)  \ln A(\nu_i)
\eeq
which is based on a Bayesian statistics, and (2) the `Laplacian',
\beq
f^{\rm Lap}(A)=\sum_i {d^2\! A(\nu)\over d\nu^2}\Bigg|_{\nu=\nu\ns_i}
\eeq
which penalizes unsmooth spectral functions (or long real-time decay). 
In  these functionals, the real frequency $\nu$ is discretized as a finite sequence $\nu_i$.
 
A different strategy is the stochastic regularization \cite{Mishchenko_Stoch,Sandvik_Stoch}. In this method the spectral function
is obtained by averaging over a large sample of randomly-chosen solutions consistent with $\chi^2/{\cal N}\approx 1$. First a random positive spectral function is generated. Then the goodness of fit is
minimized using the steepest decent method while imposing positivity at each step. This procedure is repeated until
$\chi^2/{\cal N} \approx 1$. Averaging over the random initial conditions leads to the final spectral function. 

A complementary approach is to estimate the pole structure of $A(\nu)$,  using a Pad\'{e} approximation. 
$\tilde{\CG}$  is fitted to a rational
function 
\beq
\tilde{\CG }(i\omega_m)=P_{n_p}(i\omega_m)/Q_{n_p}(i\omega_m)  ,
\eeq
 where $P_{n_p}$ and $Q_{n_p}$ are polynomials of order $n_p$. Since
$\tilde{\CG }$ is an analytic function of $i\omega_m$ we can perform the analytic continuation explicitly by taking
$\tilde{A}(\omega)=\Imag\tilde{\CG }(i\omega_m \to \omega +i0^+)$. For the best inversion, one can  increase the value of $n_p$ until $\chi^2 / \mathcal{N}\approx 1$. Further increase of $n_p$ leads to over fitting and
the appearance of spurious poles. $n_p$ needs to be determined, with similar considerations to those determining $n_{\rm svd}$.

\begin{figure}[t!]
\includegraphics[width=0.475\textwidth]{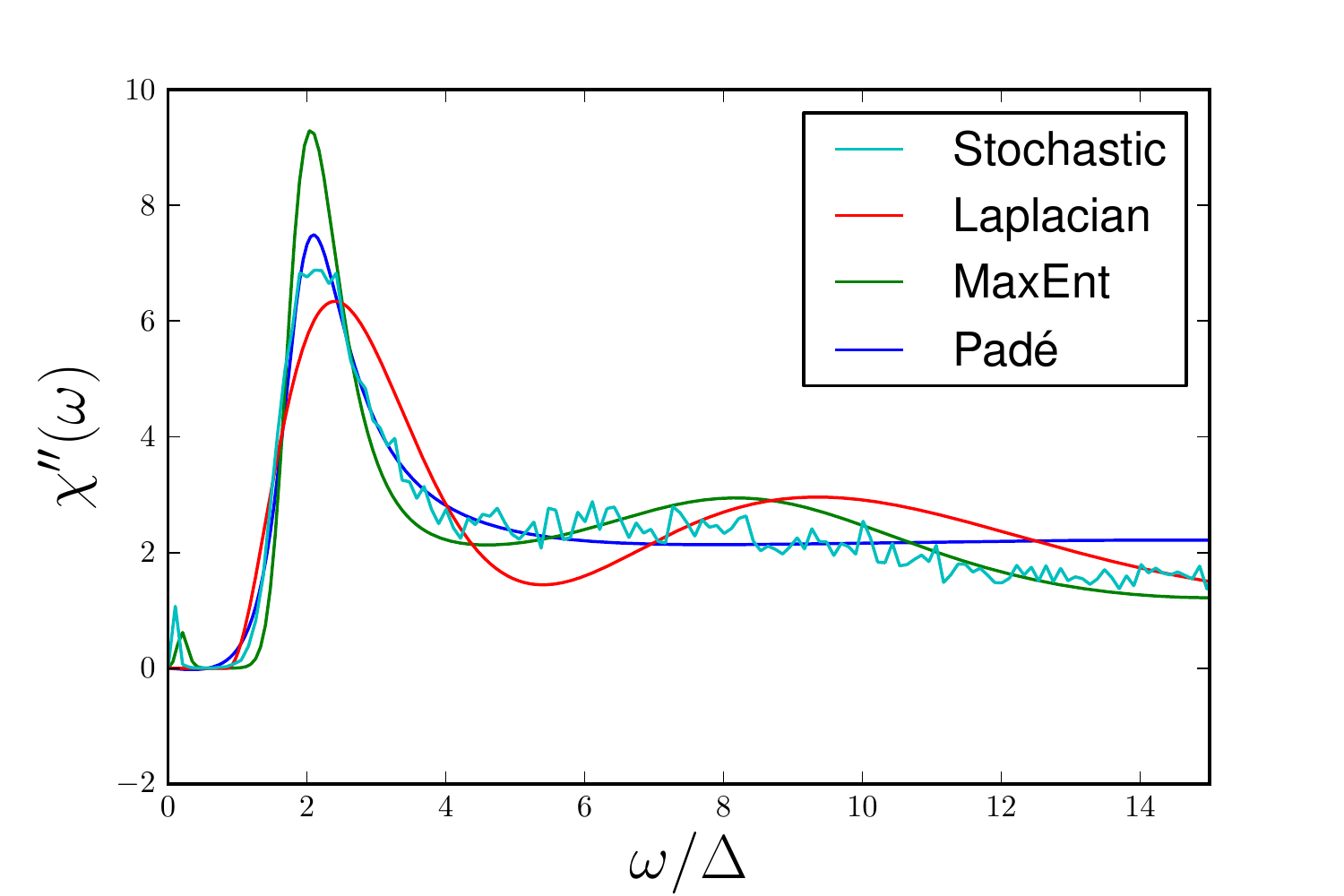}
\caption{Comparison of different regularization methods. Note that the Higgs peak position varies only slightly between the different methods, described in Section \ref{sec:MaxEnt}.}
\label{fig:compareall}
\end{figure}

In Fig.~(\ref{fig:compareall}) we show a comparison of the different regularization approaches for the same QMC data as used in Fig~(\ref{fig:svdchop}). We note
that the position of the Higgs peak varies only slightly between different analytic continuation methods, but functions differ somewhat  in the higher frequency structure. 

As a final note we comment on the form the kernel $K(i\omega ,\nu)$ for QMC simulation with discretized imaginary time axis. In this case
the imaginary time axis gets a discrete set of values $\tau_i=\Delta\tau \times i$ with $i\in\{0,\ldots,M-1\}$ with $\Delta \tau =\beta/M$.
The corresponding Matsubara frequencies are $\omega_m=2\pi n/\beta$ with $m\in\{0,\ldots,M-1\}$. The kernel is given by a sum over
all aliases of the original kernel:
\begin{equation}
\begin{split}
\tilde{K}(i\omega_m,\nu)&={1\over\pi}\sum_{k=-\infty}^\infty{2 \nu\over \big(2\pi(n+M k)/\beta\big)^{\!2}+\nu^2}\\
&=  {\beta\over M\pi}\cdot{ \sinh (\beta\nu/M)\over\cosh (\beta\nu/M) -\cos(2\pi m/M)}\ .
\end{split}
\end{equation}

\subsection{Spherical averaging}

A desirable feature of the Eq.~(\ref{eq:ContModel}) and  its discretized approximation Eq.~(\ref{eq:latModel}) is the Euclidean spacetime symmetry.  
As a consequence, it is not
necessary to single out any one specific direction as the ``time'' direction.  In particular, ignoring weak anisotropies arising from the
underlying cubic lattice, correlation functions such as that in Eq.~(\ref{eq:latSS}) are spherically symmetric and only depend on the
Euclidean distance from the point ${\bf r}=(\tau,x,y)$ to the origin.  This is especially correct near the QCP, where the large correlation
length ensures that the correlation function at long distances is insensitive to the discrete nature of the lattice.

This observation suggests that one may reduce the statistical noise by performing a spherical average over all possible time directions.
In this method, the correlation function at time $\tau$ is obtained by averaging over all the points within a thin spherical shell between
radius $r=\tau$ and $r=\tau+\delta\tau$.   This leads to a large enhancement in statistics -- for a  $L\times L\times L$ system,
${\mathcal{O}}(L^3)$ data points are used instead of the ${\mathcal{O}}(L)$ points typically used in computing the correlation function.
In order to implement this method accurately it is necessary to account for the weak anisotropy arising from the underlying cubic lattice.
This is done by projecting out the lowest cubic anisotropies prior to the averaging.  

The  bulk of the data presented in this paper  was obtained by averaging over the three principal axes only, and not taking advantage of
the full spherical averaging.   However, preliminary numerical tests show that  spherical averaging does indeed yield high quality results
while requiring shorter simulations.  This effect may be significant in light of the high sensitivity of numerical analytic continuation to
noise.  We intend to develop this strategy  further in future work.

\section{Complex conductivity}\label{AppC}

In this section we will derive the complex conductivity for the disordered and ordered phase in weak coupling.

To one loop order, the paramagnetic response in the disordered phase in given by\cite{damle_UniCond}:
\begin{equation}
\Pi_{xx}^{\rm P}(p)=\int \!\!{d^3\!q\over (2\pi)^3}\>{4 q_x^2\over q^2+\Delta^2}\cdot{1\over (q+p)^2+\Delta^2} 
\end{equation}
where $\Delta$ is the renormalized single particle gap in the disordered phase and $p=(\omega_m,0,0)$.
Introducing the Feynman parameter $x$ and shifting $q\to q-xp$,
\begin{align}
\Pi_{xx}^{\rm P}(p)=\int\limits_0^1\!\! dx\!\int\!\!{d^3 q\over(2\pi)^3} {4q_x^2\over\big[q^2+x(1-x)p^2+\Delta^2\big]^2}\ .
\end{align}
Performing the $q$ integration up to a cutoff $\Lambda$ we obtain:
\begin{equation}
\Pi_{xx}^{\rm P}(p)={2\over 3\pi^2}\int\limits_0^1 \!\! dx \> \bigg[\Lambda- {3\pi \over 4}\sqrt{p^2 x(1-x)+\Delta^2}\bigg]
\end{equation}
up to corrections that vanish as $\Lambda\to\infty$. To obtain the full response we must subtract the diamagnetic part.  Since the superfluid stiffness vanishes in the disordered phase,
this is given by $\Pi_{xx}^{\rm D}=\Pi_{xx}^{\rm P}(p\to 0)$.  This cancels the linearly divergent term, to yield
\begin{align}
\Pi_{xx}(p)&=-{1\over2\pi}\int\limits_0^1 \!\! dx \> \bigg[\sqrt{p^2x(1-x)+\Delta^2}-\Delta\bigg]\\
&= {\Delta \over 4\pi} - i {4 \Delta^2+p ^2\over 16\pi p} \ln\! \left({2 \Delta-i p\over 2 \Delta+i p} \right)\ .
\end{align}
We analytically continue by taking $p\to -i\omega+\epsilon$, resulting in
\begin{equation}
\Pi_{xx}(\omega) = {\Delta \over 4\pi} + {4 \Delta^2-\omega ^2\over 16\pi \omega} \ln\! \left({2 \Delta-\omega-i\epsilon \over 2 \Delta+\omega+i\epsilon} \right)\ .
\end{equation}
The conductivity is then
\begin{equation}
\begin{split}
\sigma(\omega)& = {1\over i\omega} \Pi_{xx}(\omega)= \Real \sigma (\omega) + i \Imag \sigma (\omega)\\
&= {1\over \omega} \Imag \Pi_{xx}(\omega) - {i\over \omega} \Real \Pi_{xx}(\omega)
\label{eq:conddis}
\end{split}
\end{equation}
We note that $\Real \sigma(\omega)$ vanishes for $\omega\ns<2\Delta$.  Above the threshold we obtain:
\begin{equation}
\Real \sigma(\omega)={\omega ^2-4\Delta^2\over 16\omega^2} \qquad (\omega>2\Delta)\ .
\label{eq:disrecond}
\end{equation}
The imaginary part is given by:
\begin{equation}
\Imag \sigma(\omega)={1\over 16\pi\omega^2}\Bigg[(4 \Delta^2-\omega ^2) \ln \left|{2 \Delta- \omega\over 2 \Delta+ \omega} \right|+4\Delta \omega\Bigg]\label{eq:disimcond}
\end{equation}

In the ordered phase the paramagnetic response is given by\cite{Lindner_BadMetal,Podolsky_visibility}:
\begin{equation}
\Pi_{xx}^{\rm P}(p)=\int \! {d^3 q\over (2\pi)^3} \> {4q_x^2\over q^2+m^2}\cdot {1\over (q+p)^2}\ .
\end{equation}
Here $m$ is the Higgs mass. As before we introduce the Feynman parameter $x$ and shift $q\to q-x p$:
\begin{align}
\Pi_{xx}^{\rm P}(p)&=\int\limits_0^1\!\! dx\!\int\!\!{d^3 q\over(2\pi)^3}\  \times\\
&\qquad {4q_x^2\over\big[q^2+(1-x)(xp^2+m^2)\big]^2}\ .\nonumber
\end{align}
Performing the $q$ integration we obtain:
\begin{equation}
\Pi_{xx}(p) =\rho_s+{m (3 m^2+5 p^2)\over 24\pi p^2}- i {(p ^2+m^2)^2\over16\pi p^3} \ln\! \left({m-i p \over m+i p }\right)\ .
\end{equation}
In the final expression we absorbed the constant term (including the linear divergence) and the diamagnetic contribution into the superfluid stiffness $\rho_s$.

After analytic continuation the real conductivity is given by:
\begin{equation}
\Real \sigma(\omega)=\pi \rho_s \delta(\omega) + {(\omega ^2-m^2)^2\over 16 \omega ^4}\> \Theta(\omega-m)\ ,
\end{equation}
with $\rho_s$ being the superfluid stiffness.
The imaginary part of the conductivity is
\begin{align}
\Imag \sigma(\omega)&={\rho_s\over\omega} +{ (m^2-\omega ^2)^2\over 16\pi\omega^4}
\ln \left|{ m-\omega\over m+\omega }\right| \nonumber \\
& \qquad +{m(3m^2-5\omega^2)\over 24\pi\omega^3}
\end{align}

\vfill\eject

\bibliography{higgsPRB}{}

\end{document}